\documentclass[10pt]{article}
\usepackage{amsmath}
\usepackage{graphicx}
\usepackage{amsfonts}
\usepackage{amssymb}
\usepackage{epsf}
\usepackage{latexsym}

\def\d{\mbox{d}}

\def\sh{\mbox{\sffamily h}}
\def\sW{\mbox{\sffamily W}}

\def\sfj{\mbox{\sffamily{\scriptsize j}}}
\def\sfl{\mbox{\sffamily{\scriptsize l}}}

\def\sM{\mbox{\scriptsize M}}

\def\sN{\mbox{\scriptsize N}}

\def\B{\mbox{\tiny B}}

\def\be{\begin{equation}}
\def\ee{\end{equation}}
\def\bea{\begin{eqnarray}}
\def\eea{\end{eqnarray}}
\def\fn{\footnote}
\def\pa{\partial}
\def\d{\textrm{d}}

\def\Di{\textrm{d}_{\underline{\mbox{\scriptsize B}}}}

\def\DotB{\mbox{\Large$\circ$}_{\underline{{\mbox{\scriptsize B}}}}}

\def\hat{\widehat}
\def\Kappa{\mbox{\Large $\kappa$}}
\def\Gp{\Gamma^{\prime}}
\def\Dp{\Delta^{\prime}}

\def\Gpp{\Gamma^{\prime\prime}}
\def\Dpp{\Delta^{\prime\prime}}

\def\B{\mbox{\underline{B}}}
\def\suB{\mbox{\scriptsize\underline{B}}}

\def\P{\mbox{\underline{P}}}

\def\nj{\mbox{{j}}}
\def\nl{\mbox{{l}}}

\def\nn{\mbox{{j}}}
\def\np{\mbox{{p}}}
\def\nB{\mbox{{B}}}
\def\nH{\mbox{{H}}}
\def\nL{\mbox{{L}}}
\def\nP{\mbox{{P}}}
\def\nR{\mbox{{R}}}

\def\sa{\mbox{\scriptsize a}}

\def\scr{\mbox{\scriptsize (cross)}}
\def\sd{\mbox{\scriptsize d}}
\def\sg{\mbox{\scriptsize g}}
\def\sh{\mbox{\scriptsize h}} 
\def\sj{\mbox{\scriptsize j}} 
\def\sll{\mbox{\scriptsize l}} 
\def\sm{\mbox{\scriptsize m}}
\def\smom{\mbox{\scriptsize mom}}

\def\sn{\mbox{\scriptsize n}} 
\def\sp{\mbox{\scriptsize p}} 
 
\def\sw{\mbox{\scriptsize w}}
\def\sA{\mbox{\scriptsize A}}

\def\sY{\mbox{\scriptsize Y}}
\def\sBO{\mbox{\scriptsize BO}}
\def\sAM{\mbox{\scriptsize AM}}
\def\sPl{\mbox{\scriptsize Pl}}
\def\sBOB{\mbox{\scriptsize BOB}}

\def\sH{\mbox{\scriptsize H}}
\def\sI {\mbox{\scriptsize I}}

\def\sL{\mbox{\scriptsize L}}

\def\sR{\mbox{\scriptsize R}}

\def\sT{\mbox{\scriptsize T}}
\def\sV{\mbox{\scriptsize V}}
\def\sW{\mbox{\scriptsize W}}

\def\Pl{\mbox{\scriptsize Pl}}
\def\scosmic{\mbox{\scriptsize cosmic}}
\def\eph{\mbox{\scriptsize eph}}
 
\def\eph(B){\mbox{\scriptsize emergent(LMB)}} 
\def\sem{\mbox{\scriptsize em}} 
\def\seB{\mbox{\scriptsize emergent(LMB)}} 
\def\TL{T^{\mbox{\scriptsize emergent(LMB)}}}
\def\tTL{T^{\mbox{\tiny emergent(LMB)}}}
\def\T0L{T^{\mbox{\scriptsize emergent(LMB)}}(0)}
\def\TaL{T^{\mbox{\scriptsize emergent(LMB)}}_{0}  }

\def\sWKB{\mbox{\scriptsize emergent(WKB)}}
\def\ssWKB{\mbox{\scriptsize WKB}}
\def\tWKB{\mbox{\tiny WKB}}
\def\WKBH{\mbox{\scriptsize emergent(WKB: H)}}
\def\WKBL{\mbox{\scriptsize emergent(WKB: L)}}

\def\sYo{\mbox{\scriptsize internal(York)}}

\def\semi{\mbox{\scriptsize emergent(WKB-LMB)}}

\def\th{\mbox{\tiny h}}

\def\ttH{\mbox{\tt{H}}}

\def\fj{\mbox{\sffamily j}}
\def\fl{\mbox{\sffamily l}}

\def\fF{\mbox{\sffamily F}}

\def\fS{\mbox{\sffamily S}}
\def\fT{\mbox{\sffamily T}}

\def\fW{\mbox{\sffamily W}}
\def\aha{\mbox{N}}
\begin{document}

\begin{titlepage}

\begin{center}

\vspace{.3in}

{\Large{\bf EMERGENT SEMICLASSICAL TIME IN QUANTUM GRAVITY. }} 

\vspace{.1in}

{\large{\bf II. FULL GEOMETRODYNAMICS AND MINISUPERSPACE EXAMPLES}} 

\vspace{.3in}

{\large{\bf Edward Anderson}}

\vspace{.3in}

\noindent{\em Peterhouse, Cambridge, U.K., CB21RD;}

\noindent{\em DAMTP, Centre for Mathemetical Sciences, Wilberforce Road, Cambridge, U.K., CB30WA.}

\end{center}

\vspace{.3in}

%\baselineskip=24pt 

%===========================================================ABSTRACT=======================================================================
\begin{abstract}

\noindent I apply the preceding paper's semiclassical treatment to geometrodynamics.
The analogy between the two papers is quite useful at the level of the quadratic 
constraints, while I document the differences between the two due to the underlying 
differences in their linear constraints.  
I provide a specific minisuperspace example for my emergent semiclassical time scheme and 
compare it with the hidden York time scheme.  
Overall, interesting connections are shown between Newtonian, Leibniz--Mach--Barbour, WKB 
and cosmic times, while the Euler and York hidden dilational times are argued to be 
somewhat different from these.
\end{abstract}

%==========================================================================================================================================

\vspace{.3in}

\mbox{ }

%===============================================COMMENTS FOR PREPRINT ARCHIVE==============================================================

%\noindent{\bf Keywords:} 

%==========================================================================================================================================

\noindent{\bf PACS numbers 04.60-m, 04.60.Ds}

\end{titlepage}

%=====================================================================================================
%=====================================================================================================
%==========================================================================================================================================
\section{Introduction}
%==========================================================================================================================================
%==========================================================================================================================================
%==========================================================================================================================================

This Paper is the geometrodynamical sequel of the preceding relational particle model (RPM) paper \cite{TFNS}. 
It considers emergent and hidden timefunction approaches to the problem of time in quantum gravity 
\cite{Kuchar92, Isham93, Wheeler, DeWitt, EarlyKuchar, Kuchar81, UW89, K91, B94II, EOT, Kuchar99, 
Kieferbook, OII, TLESS}.  
Let the configurations now be fields $\theta_{\Gamma}(x)$ (for $\Gamma$ indexing both 
field species and the spatial, internal indices of each field), taken to include 
the 3-metric $h_{\alpha\beta}$.  
Let these have kinetic term ${\cal T}_{\theta} = _{\underline{{\Theta}}^{-1}}||\DotB\theta||^2$, 
taken to be homogeneous quadratic in their velocities, where $\mbox{}_\Theta|| \mbox{ } ||$ 
is the norm with respect to the undensitized and now generally nondiagonal and configuration-dependent 
array ${\underline{\Theta}}^{\Gamma\Delta}(x;\theta_{\Sigma}(x)]$: the configuration 
space metric.  
Nor is this now necessarily positive-definite (while I steer 
free of it being degenerate or velocity-dependent).  
See Paper I's Appendix A for the $\DotB$ symbol, 
and its footnote 1 for the rest of the notation in common.
The fields' potential is ${\cal V}_{\theta} \equiv {\cal V}(x;\theta_{\Sigma}(x)] \equiv - {\cal U}_{\theta}$. 
The action is now
\be
\fS_{\mbox{\scriptsize BSW-type}}[\theta_{\Gamma}, \dot{\theta}_{\Gamma}, \dot{\nB}_{\alpha}] = 
\int \d\lambda \int \d^3x\sqrt{h}
\sqrt{  {\cal U}_{\theta}  {\cal T}_{\theta} } \mbox{ } .
\ee
While this relational Baierlein--Sharp--Wheeler \cite{BSW, GRT} type local square root action
is not general enough to encompass all theories \cite{RWR, San, Than, Lan, Phil}, 
it does encompass general relativity (GR)\fn{
%%%%%%%%%%%%%%%%%%%%%%%%%%%%%%%%%%%%%%%%%%%%%%%%%%%%%%%%%%%%%%%%%%%%%%%%%%%%%%%%%%%%%%%%%%%%%%%%%%%%%%%%
GR, moereover, has a restricted meaning here: the geometrodynamics underlied by some fixed, 
compact without boundary spatial topology.  For discussion and justification of this, see \cite{Phil}.} 
%%%%%%%%%%%%%%%%%%%%%%%%%%%%%%%%%%%%%%%%%%%%%%%%%%%%%%%%%%%%%%%%%%%%%%%%%%%%%%%%%%%%%%%%%%%%%%%%%%%%%%%%
(see \cite{BSW, B94I, RWR, ABFKO}, or Sec I.1) alongside conventional minimally-coupled fundamental bosonic 
matter fields \cite{RWR, AB, Van, Than, Lan}.  
This suffices for the present study (and the extension which includes sufficient conventional fields 
of all spins, fermionic as well as bosonic, is straightforward and only slightly more cumbersome 
\cite{Van}). 
See \cite{reltech} for related techniques.

The conjugate momenta are:  
\be
\Pi^{\Gamma}(x) \equiv \frac{\delta \fS}{\delta \dot{\theta}_{\Gamma}} = 
\sqrt{              \frac{{\cal U}_{\theta}         }
                                                   {{\cal T}_{\theta}         }              }
\Theta^{\Gamma\Delta}
%[\theta_{\Sigma}(x); x)
\DotB\theta_{\Delta} 
\mbox{ } \mbox{ . } \mbox{ } 
\ee
Then, working along the same lines as in Sec I.1, the reparametrization invariance which implements 
the temporal relationalism leads to a primary constraint  
\be
\mbox{Quad}(x;\theta_{\Sigma}, \Pi^{\Sigma}] = 
\mbox{ }_{{{\Theta}}}||\Pi||^2 
+ \sqrt{h}{\cal V}_{\theta} = 0 \mbox{ } .  
\ee
Moreover, variation with respect to the spatial relationalism implementing 3-diffeomorphism auxiliary 
vector $\B$ gives a secondary constraint of form 
\be
\stackrel{\mbox{\Huge$\longrightarrow$}}
         {\mbox{Lin}_{\Gamma}(x;\theta_{\Sigma}]}\Pi^{\Sigma} = 0 \mbox{ }
\ee
(possibly only modulo further matter constraints, as happens when gauge theory matter is considered).  
With modern quantum cosmology \cite{BI75, HH, Hawking84, Hawking88, HallHaw, +QCos, ++QCos, K00} 
and inflation \cite{InflationBG, InflationL, InflationM, Inflation}  in mind, I specialize the above working explicitly to the case of the  
Einstein--MCMSF (minimially-coupled multi-scalar field) system.

I consider a particularly useful H--L split for this in Sec 2. 
In Sec 3 I consider the Leibniz--Mach--Barbour (LMB) emergent time notion in GR, $T^{\seB}$.
In Secs 4--8 I consider the emergent semiclassical time notion in GR, $T^{\sWKB}$  
Again, $T^{\seB}$ and $T^{\sWKB}$  are found to be in very close parallel.  
I extend Paper I's geometric approach with its detailed list of, 
and cross-checks between, approximations to 
geometrodynamics with MCMSF matter, 
considering the equations at the quantum level in Sec 4, 
the Born--Oppenheimer (BO) ansatz in Sec 5, 
and the WKB ansatz in Sec 6.  
For earlier literature on the semiclassical approach in quantum gravity/cosmology, see
\cite{DeWitt, Peres, Wheeler, semicl, LR79, Banks, HallHaw, Zehbook, Pad, PadSingh, 
KS, BV89, Kiefer94short, K00, Datta, B93, Kuchar92, Isham93, CG, Kieferbook, OII}.
In Sec 7 I consider my iterative scheme of Paper I in this new setting, 
and update my answers and discussions as regards the basic (`B') and detailed (`D') 
questions in Sec I.1.  
I provide a specific minisuperspace example in Sec 8. 
In Sec 9, I contrast this with the well-known hidden York time \cite{York72, Kuchar92, Isham93}, 
which is directly analogous to the previous paper's hidden `Euler time'.
I conclude in Sec 10.

%=====================================================================================================
%=====================================================================================================
%==========================================================================================================================================
\section{A particularly useful H-L split in geometrodynamics}
%==========================================================================================================================================
%==========================================================================================================================================
%==========================================================================================================================================

Even for the Einstein-MCMSF system, there are many ways in which 
one could identify the H's and L's among the $h_{\alpha\beta}$ and $\phi_{\Gpp}$.  
That depends partly on the intended application -- 
explaining classicality today, studying simple features of early universes, 
or studying much finer anisotropic and inhomogeneous features of early universe 
models with the origin of galaxies or the detailed structure of the CMB in mind.  
Indeed, some of these applications could involve {\sl multiple} hierarchies.
One reasonable first choice for a semiclassical study of the universe 
is to consider gravitation to be associated with the Planck mass $M_{\Pl}$, while the scalar fields 
have mass terms a number of orders of magnitude smaller than this. 
This may serve to study simple early-universe features and perhaps to 
explain late-time classicality, and as a prerequisite for the more ambitious 
goal of studying the much finer features that rest on inhomogeneities.  
A benefit of this choice alongside the choice of minimally-coupled matter, 
which is {\sl not} shared by other choices, 
is the great simplification through the kinetic 
metrics then depending on the H d.o.f.'s alone, 
in the minimally-coupled case, rendering the advantages of Sec I.5 applicable here.

Overall, the above choice enables the rewrite
\be
\Theta_{\Gamma\Delta}(x; \theta_{\Sigma}] \longrightarrow {G}_{\Gamma\Delta}(\nH_{\Gp}) = 
{\cal G}_{\alpha\beta\gamma\delta}(h_{\mu\nu}) \bigoplus \Xi_{\Gpp\Dpp}(h_{\mu\nu})
\ee
for ${\cal G}$ the DeWitt supermetric of GR, $\frac{1}{\sqrt{h}}
\left\{ h_{\alpha\gamma}h_{\beta\delta} - \frac{1}{2}h_{\alpha\beta}h_{\gamma\delta} \right\}$ 
and $\Xi$ the matching densitization of the minimally-coupled matter configuration space metric.  
Thus the Einstein--MCMSF extension of the BSW-type action is
\be
\fS_{\mbox{\scriptsize BSW-type}}
[h_{\alpha\beta}, \phi_{\Gamma}, \dot{h}_{\alpha\beta}, \dot{\phi}_{\Gamma}, \dot{\nB}_{\alpha} ] = 
\int \d\lambda \int \d^3x\sqrt{h}
\sqrt{        
\left\{    
{\cal U}_{\sh} + {\cal U}_{\phi} + {\cal J}_{\sh\phi}  
\right\}
\left\{    
{\cal T}_{\sh} + {\cal T}_{\phi}
\right\}      } 
\mbox{ } ,
\label{leia}
\ee
where (dropping the double primes from now on)
\be
{\cal T}_{\sh} = {\cal T}_{\sh}(x^{\mu}, h_{\alpha\beta}, \dot{h}_{\alpha\beta}; \dot{\nB}_{\alpha}] = 
\mbox{}_{\underline{\sM}}||\DotB h||^2  
\mbox{ } ,\mbox{ } 
\underline{M}^{\alpha\beta\gamma\delta} = M_{\Pl}^2 \frac{c}{\hbar}\{{\cal G}^{-1}\}^{\alpha\beta\gamma\delta} \mbox{ } , 
\ee
\be
{\cal T}_{\phi} = {\cal T}_{\phi}(x^{\mu}, \dot{\phi}_{\Gamma}; \dot{\nB}_{\alpha}] = 
                            \mbox{}_{\sm}||\DotB\phi ||^2 
\mbox{ } ,\mbox{ } \underline{m}^{\Gamma\Delta} \mbox{ going as } 
m^2\frac{c}{\hbar}\{\underline{\Phi}^{-1}\}^{\Gamma\Delta}
\ee
is the kinetic mass matrix for the scalar fields for $\Phi^{-1}$ a dimensionless mass ratio matrix.  
m is a `representative mass', so 
\be
\stackrel{\mbox{max}}{\Gamma, \Delta}\frac{|m_{\Gamma} - m_{\Delta}|}{m_{\Gamma}} = \varepsilon_{\Delta m} \mbox{ } , \mbox{ small } .
\ee
I also assume that
\be
{m^2}/{M^2_{\mbox{\scriptsize Pl}}} = \varepsilon_{\sH\sL} \mbox{ , small .}
\ee
Thus, overall I assume a sharply-peaked hierarchy.\fn{How useful are these assumptions? 
As regards the inflaton field, 
$m_{\phi}/M_{\Pl} = 10^{-5}$ or $10^{-6}$ from observations of inhomogeneities on the galactic scale 
(see \cite{InflationL} or \cite{InflationM}).  
On the other hand, multi-scalar field models 
could well not have a simplifying sharply-peaked mass hierarchy.}
Note the $M_{\sH}$--$M^2_{\Pl}$ analogy between the 2 papers.
Also, 
\be
{\cal U}_{\sh}(x^{\mu}; h_{\alpha\beta}] = \frac{M_{\Pl}c^3}{\hbar}\{{\cal R} + \Lambda\} 
\equiv \tilde{\cal R} + \tilde{\Lambda}
\mbox{ } , \mbox{ }
\ee
\be
{\cal U}_{\phi}(\phi_{\Gamma}) = - {\cal V}_{\phi}(\phi_{\Gamma}) = 
- \frac{c^3}{\hbar} \sum_{\Gamma}\frac{m^2_{\Gamma}}{2}\omega_{\Gamma}^2
\phi_{\Gamma}^2 - {\cal V}_{\phi}^{\mbox{\scriptsize int}}
\mbox{ } , \mbox{ for } \mbox{ } 
\omega_{\Gamma} = \frac{m_{\Gamma}c^2}{\hbar} \mbox{ the `Einstein--Planck' frequency } 
\mbox{ } , \mbox{ }
\ee
\be
{\cal J}_{\sh\phi}(x^{\mu}, h_{\alpha\beta}; \phi_{\Gamma}] = -
{\cal I}_{\sh\phi}(x^{\mu}, h_{\alpha\beta}; \phi_{\Gamma}] 
                   = h^{\alpha\beta}{\cal Q}^{\Gamma\Delta}\lfloor 
\pa_{\alpha} \phi_{\Gamma} \rfloor \pa_{\beta}\phi_{\Delta} /4
\mbox{ } , \mbox{ where } \mbox{ }
{\cal Q} = \frac{m^2c^3}{\hbar}Q^{\Gamma\Delta} 
\mbox{ } ,  
\ee
for $Q^{\Gamma\Delta}$ a dimensionless matrix.

The corresponding (redundant) 
configuration space is {\bf Riem} $\times {\mbox{\bf$\Phi$}} \times$ {\bf B}, 
for {\bf Riem} the space of spatial 3-metrics on a fixed topology taken here to be one which is 
compact and without boundary (and equipped pointwise with the DeWitt supermetric), 
$\mbox{\bf $\Phi$}$ is the space of scalar fields 
(equipped with the constant kinetic and potential metrics $\Phi$, ${\cal Q}$), and {\bf B} is the 
space of the auxiliary fields $\nB_{\alpha}$.  
The geometrodynamical and MCMSF momenta are then,   
\be
\pi^{\alpha\beta} = \sqrt{\frac{{\cal U}_{\sh} + {\cal U}_{\phi} + {\cal J}_{\sh\phi}} 
                               {{\cal T}_{\sh} + {\cal T}_{\phi}}}
{M}^{\alpha\beta\gamma\delta}\DotB h_{\gamma\delta} 
\mbox{ } \mbox{ , } \mbox{ } 
\np^{\Gamma} = \sqrt{\frac{{\cal U}_{\sh} + {\cal U}_{\phi}  +  {\cal J}_{\sh\phi}}
                          {{\cal T}_{\sh} + {\cal T}_{\phi}}}{m}^{\Gamma\Delta}\DotB\phi_{\Delta} \mbox{ } ,
\label{sh}
\ee
while the above constraint working produces 
\be
\hat{\cal H} = 
\mbox{}_{\sM^{-1}}||\pi_{\sh}||^2 + \mbox{ }_{\sm^{-1}}||p_{\phi}||^2 - 
\sqrt{h}\frac{c^4}{16\pi G}
\{{\cal R} + \Lambda \} 
+ \sqrt{h}\{  {\cal V}_{\phi} + {\cal I}_{\sh\phi} \}  = 0 \mbox{ }        
\mbox{ (Einstein--MCMSF Hamiltonian constraint) } ,
\ee
\be
- 2\nabla_{\beta}\pi^{\alpha\beta} + {p_{\phi}}^{\Gamma}\pa^{\alpha}\phi_{\Gamma} = 0      \mbox{ (Einstein--MCMSF momentum constraint) }
\mbox{ } .
\ee

%=====================================================================================================
%=====================================================================================================
\section{Leibniz--Mach--Barbour (LMB) time in geometrodynamics}
%=====================================================================================================
%=====================================================================================================

Sec. 1--2's formulation for geometrodynamics based on the action (\ref{leia}) 
or, with reparametrization invariance made explicit,

\noindent
\be
\fS[h_{\alpha\beta}, \phi_{\Gamma}, \Di h_{\alpha\beta}, \Di{\phi}_{\Gamma}, \nB_{\alpha} ] = 
\int  \int \d^3x\sqrt{h}
\sqrt{        
\left\{    
{\cal U}_{\sh} + {\cal U}_{\phi} + {\cal J}_{\sh\phi}     
\right\}
\left\{    
\mbox{}_{\underline{\sM}}    ||\Di h_{\alpha\beta}(x)||^2    +      
\mbox{}_{\underline{\sm}}       ||\Di\phi_{\Gamma}||^2     
\right\}      } 
\mbox{ } ,
\ee 
can be interpreted as having an emergent quantity 

\noindent
\be
\frac{S}{2}\sqrt{          \frac{       {\cal T}_{\sh} + {\cal T}_{\phi}                                }        
                     {     {\cal U}_{\sh} + {\cal U}_{\phi} + {\cal J}_{\sh\phi}       }          }
= \mbox{N}(x^{\mu}, \dot{h}_{\alpha\beta}, \dot{\phi}_{\Gamma}; h_{\alpha\beta}, \phi_{\Gamma}]
\ee
(up to a constant scale $S$) 
in its momentum-velocity relations (\ref{sh}) and in its Euler--Lagrange equations (not provided).  
Because it is dimensionally a velocity, I also on occasion denote it by $\dot{\mbox{A}}$.  
Furthermore, one can interpret the particular combination 
$\frac{1}{\mbox{\sN}}\frac{\pa}{\pa\lambda} $ which occurs in the abovementioned equations 
as $\frac{\pa}{\pa \tTL}$ for $\TL(x^{\mu}; h_{\alpha\beta}, \phi_{\Gamma}]$ 
the {\it LMB time of GR} \cite{B94I}.
Then, explicitly, integrating,
\be
\TL(x^{\mu}; h_{\alpha\beta}, \phi_{\Gamma}] = 
\T0L + 
\frac{S}{2}\int
\sqrt{          \frac{      \mbox{}_{\underline{\sM}}  
                                               ||\Di h_{\alpha\beta}(x)||^2  
                            + \mbox{}_{\underline{\sm}      }||\Di \phi_{\Gamma}||^2        }
                     {      {\cal U}_{\sh} + 
                            {\cal U}_{\phi} + {\cal J}_{\sh\phi}     }         } 
\mbox{ } .
\label{TL}
\ee 
This is a classical time, and is provided by the system itself rather than being an external time.  
It is a measure of change in the `whole'\fn{
%%%%%%%%%%%%%%%%%%%%%%%%%%%%%%%%%%%%%%%%%%%%%%%%%%%%%%%%%%%%%%%%%%%%%%%%%%%%%%%%%%%%%%%%%%%%%%%%%%%%%%%%
As in Paper I, this totality excludes linear kinetic terms, were fields with such incorporated (which 
now includes the important example of Einstein--Dirac theory).}  
%%%%%%%%%%%%%%%%%%%%%%%%%%%%%%%%%%%%%%%%%%%%%%%%%%%%%%%%%%%%%
configuration.  
The $\TL(0)$ term here plays the familiar role of choice of time-origin.  
N.B. that unlike for the RPM, the lapse and LMB times of GR are locally defined, i.e. 
in general vary from spatial point to spatial point.  
From now on, the scale choice S = 1 is in use.    
Then $\mbox{N} \equiv \mbox{N}_1$ is the Arnowitt--Deser--Misner \cite{ADM} lapse of GR.  
$\dot{\mbox{A}}$ $\equiv \dot{\mbox{A}}_1$.  $\ast \equiv \ast_1$

\noindent\underline{Classical time lemma of GR cosmology}
Choosing constant C = N amounts to picking the label time to be $T^{\seB}$ up to origin and scale.  
This choice simplifies equations and is equivalent to the choice of cosmic time, $T^{\scosmic}$.  

\noindent The first statement is proven in direct analogy with its Paper I counterpart.  
Moreover, N $= 1$ is the cosmic time partial gauge fixing of GR (also known in a wider context 
as the synchronous partial gauge fixing).

\noindent   
$T^{\scosmic}$ is particularly clearly defined and understood in the case of homogeneous cosmologies.

How does $\TL$ fare as regards Sec I.1's three desirable properties of wavefunctions?  

\noindent \underline{1: globality.} As it does not necessarily exist at zeros of the denominator
${\cal U}_{\sh} + {\cal U}_{\phi} + {\cal J}_{\sh\phi}$, 
it is not generally globally valid for a given geometrodynamical motion.
Although, there is a significant 
difference with Paper I in the numerator being indefinite - there is more scope for the regions on both 
sides of a zero to have real emergent time.  
The zeros sometimes correspond to points at which $\fS$ goes complex.  
Indeed the action $\fS = \int \d\lambda\int \d^3x\sqrt{h}
\sqrt{   \{   {\cal U}_{\sh} + {\cal U}_{\phi} +{\cal J}_{\sh\phi}   \}
         \{   {\cal T}_{\sh} + {\cal T}_{\phi}   \}}$ itself may 
well cease to make sense at such zeros, through itself becoming complex.
These zeros are not now in general `halting points' in the sense of $\P_{i} = 0$ there: 
$0 = {\cal U}_{\sh} + {\cal U}_{\phi} + {\cal J}_{\sh\phi} = \fT(\P_i)$ 
by conservation of energy but now $\fT$ is not positive-definite.  
But overall, $\TL$ will not always serve as a global timefunction for geometrodynamical motions.  
It may sometimes be possible to redefine the timestandard to move past such zeros, in some cases 
obtaining a fuller range of real values and in other cases as an analytic continuation into the 
complex plane.  
Complex action, momentum, time correspond to classically forbidden regions, but these can play 
a QM role (through being penetrated by decaying wavefunctions).  
Nor is the synchronous gauge globally well-lived in general.  

\noindent \underline{2: monotonicity.}  However, if N exists as a real function for (a given portion of) 
a given motion, the monotonicity of $\TL$ is 
guaranteed thereupon: $\pa  T^{\eph(B)}/{\pa\lambda}  \geq 0$.
[Note this is not a $\lambda$-dependent statement by `cancellation' -- is invariant under 
the valid reparametizations of $\lambda$ since these themselves are monotonic.]  
While existence is not compromised by sufficiently benign blow-ups in N,  
i.e . those for which it remains integrable, such a blowup corresponds to the $\TL$ graph 
becoming infinite in slope.   
There may also be frozenness: at points for which the graph is horizontal, i.e. 
${\cal T} = 0$ or ${\cal U}_{\sh} + {\cal U}_{\phi} + {\cal J}_{\sh\phi} $ infinite.
Both zero and infinite slope may compromise use of $\TL$ itself to keep track 
for some ranges of geometrodynamical motion.  
But at least in some cases, redefined timestandards may permit the following of 
motions through such points.

\noindent \underline{3: operational meaningfulness.} The problems with observing $T^{\seB}$ itself,  
or with using more readily observable 
approximations to it, would be expected to carry over to the present geometrodynamical case.  
%page* edit 
%

%====================================================================================================
\noindent\underline{Further analysis in the H-L split regime.}
%====================================================================================================
%
%
%
The expression (\ref{TL}) becomes 
\be
\TL - \T0L = \TaL
\left\{
1 + O
\left(
\varepsilon_{\sL\sH}; \varepsilon_{\sV}, \varepsilon_{\sI}, \varepsilon_{\sT} 
\right]
\right\}
\label{hint}
\ee
for
\be
\TaL = \frac{1}{2c}\int    \mbox{}_{\underline{\cal G}^{-1}}||\Di h_{\alpha\beta}||
                          /\sqrt{\Lambda + {\cal R}} \mbox{ } , 
\label{mint}
\ee
and assuming that 
\be
\left|
{\cal V}_{\phi}/{\cal V}_{\sh}
\right| 
= \varepsilon_{\sV} 
\mbox{ }  \mbox{ ($\phi$-potential subdominance) } ,
\label{epsV}
\ee
\be
\left|
{\cal I}_{\sh\phi}/{\cal V}_{\sh}
\right| 
= \varepsilon_{\sI} 
\mbox{ } \mbox{ (interaction potential subdominance)} , 
\label{epsI}
\ee
and 
\be 
\left|
{\cal T}_{\phi}/{\cal T}_{\sh}
\right| = \epsilon_{\sT}  
\mbox{ } \mbox{ ($\phi$-kinetic subdominance) }\mbox{ }    
\label{epsT} 
\ee
are all small. 
Evaluating (\ref{mint}) provides an approximate LMB time standard.  
Critiques of the types raised in I.3.2-3 should also be heeded here.

%==========================================================================================================================================
%==========================================================================================================================================
%=====================================================================================================
%\macrosection{The semiclassical approach}
%=====================================================================================================
%==========================================================================================================================================
%==========================================================================================================================================

%=====================================================================================================
%=====================================================================================================
\section{Quantized H-L split geometrodynamics}
%=====================================================================================================
%=====================================================================================================

The following diagram commutes (albeit with operator ordering ambiguities in whichever  
passage to the last row).  
$$
\hspace{1.2in}
\fS(\dot{\theta}_{\Gamma}; \theta_{\Gamma}, \dot{\nB}_{\alpha}] 
\hspace{0.5in}
\stackrel{\mbox{\scriptsize H--L split}}{\longrightarrow} 
\hspace{0.5in}
\fS_{\mbox{\scriptsize BSW-type}}[ \dot{h}_{\alpha\beta} , 
\dot{\phi}_{\Gamma}; h_{\alpha\beta} , \phi_{\Gamma},  \dot{\nB}_{\alpha}] \hspace{0.7in}
$$
$$
\stackrel{\mbox{\scriptsize variation, inspection}}{\mbox{\scriptsize of momenta}} 
\downarrow 
\hspace{2in} 
\downarrow
\stackrel{\mbox{\scriptsize variation, inspection}}{\mbox{\scriptsize of momenta}} \hspace{0.9in}
$$
$$
\hspace{0.4in}
\left(
\stackrel{\mbox{${\cal H}$}}{\mbox{${\cal M}^{\alpha}$}}
\right)
(x^{\mu}; \theta_{\Gamma}, \Pi^{\Gamma}] = 
\left(
\stackrel{\mbox{$\sqrt{h}\Lambda$}}{\mbox{$0$}}
\right)
\hspace{0.7in}
\stackrel{\mbox{\scriptsize H--L split}}{\longrightarrow} 
\hspace{0.5in}
\left(
\stackrel{\mbox{${\cal H}$}}{\mbox{${\cal M}^{\alpha}$}}
\right)
(x^{\mu}; h_{\alpha\beta}, \phi_{\Gamma}, \pi^{\alpha\beta}, \np^{\Gamma}] = 
\left(
\stackrel{\mbox{$\sqrt{h}\Lambda$}}{\mbox{$0$}}
\right)         \hspace{0.55in}
$$
$$
\stackrel{        \mbox{\scriptsize position representation quantization}        }
         {        \mbox{\scriptsize$( \nR_{\Gamma}, \nP^{\Gamma} ) \mapsto 
          (\hat{\theta}_{\Gamma}, \hat{\Pi}^{\Gamma} ) = 
          (\theta_{\Gamma}, -i\hbar{\delta_{\theta}}^{\Gamma})$ }        } 
\hspace{0.2in}
\downarrow 
\hspace{1.95in}
\downarrow 
\stackrel{     \mbox{\scriptsize position representation quantization 
                $(h_{\alpha\beta}, \phi_{\Gamma}, \pi^{\alpha\beta}, \np^{\Gamma})$}     } 
         {     \mbox{\scriptsize $\mapsto 
(\hat{h}_{\alpha\beta}, \hat{\phi}_{\Gamma}, \hat{\pi}^{\alpha\beta}, \hat{\np}^{\Gamma})   
  =    (h_{\alpha\beta}, \phi_{\Gamma} , -i\hbar{{\delta}_{\th}}^{\alpha\beta}, 
                                        -i\hbar{{\delta}_{\phi}}^{\Gamma}   )$}     }        
\hspace{0.2in}
$$
\be
\hspace{0.8in}
\left(
\stackrel{    \mbox{$\hat{{\cal H}}$}    }{   \mbox{$\hat{{\cal M}}^{\alpha}$}    }
\right)
(x^{\mu}; \theta_{\Gamma}, {\delta_{\sR}}^{\Gamma}] = 
\left(
\stackrel{\mbox{$\sqrt{h}\Lambda$}}{\mbox{$0$}}
\right)
\hspace{0.5in}
\stackrel{    \mbox{\scriptsize H--L split}    }{    \longrightarrow    } 
\hspace{0.5in}
\left(
\stackrel{    \mbox{$\hat{{\cal H}}$}    }{   \mbox{$\hat{{\cal M}}^{\alpha}$}    }
\right)
(x^{\mu}; h_{\alpha\beta}, \phi_{\Gamma} , {{\delta}_{\sh}}^{\alpha\beta}, {{\delta}_{\phi}}^{\Gamma}  ] = 
\left(
\stackrel{\mbox{$\sqrt{h}\Lambda$}}{\mbox{$0$}}
\right) \mbox{ } . \hspace{0.1in}
\ee
So, by whichever path, the quantum energy constraint is 
\be
\hat{{\cal H}}\Psi = \hat{{\cal H}}_{\sh}\Psi + \hat{{\cal H}}_{\sh\phi}\Psi = \sqrt{h}\Lambda\Psi 
\mbox{ } ,
\ee
for
\be
\hat{{\cal H}}_{\sh\phi} \equiv \hat{{\cal H}}_{\phi} + {\cal I}_{\sh\phi} 
\mbox{ } , \mbox{ }   
\hat{{\cal H}}_{\sh} \equiv - \hbar^2\mbox{}_{\sM^{-1}}||\delta_{\sh}||^2 + {\cal V}_{\sh}
\mbox{ } , \mbox{ }
\hat{{\cal H}}_{\phi} \equiv - \hbar^2\mbox{}_{\sm^{-1}}||\delta_{\phi}||^2 + 
{\cal V}_{\phi} 
\mbox{ } . \hspace{0.05 in}
\label{full}
\ee
Moreover, the quantum momentum constraint is
\be
\hat{{\cal M}}^{\alpha}\Psi = 
\hat{{\cal M}}_{\sh}\mbox{}^{\alpha}\Psi + 
\hat{{\cal M}}_{\phi}\mbox{}^{\alpha}\Psi = 0
\label{QAM} 
\mbox{ } ,
\ee
for 
\be
{\hat{{\cal M}}_{\sh}}\mbox{}^{\alpha} = -2\frac{\hbar}{i}\nabla_{\beta}{\delta_{\sh}}^{\alpha\beta}
\mbox{ } , \mbox{ } 
\hat{{\cal M}}_{\phi}\mbox{}^{\alpha} = \frac{\hbar}{i}\lfloor\pa^{\alpha}\phi_{\Gamma}\rfloor{\delta_{\phi}}^{\Gamma} \mbox{ } .  
\label{HLAMQ}
\ee
I next lay down the standard semiclassical approach ans\"{a}tze and approximations, 
alongside objections to using these in as the present closed universe context.    
I form `less approximate' equations first, to make it clear which further approximations 
are required to go between these and more standard, more approximate forms, and also to keep explicit  
track of these smallnesses and any inter-relations between them.

%====================================================================================================
%====================================================================================================
\section{The Born--Oppenheimer (BO) type  scheme} 
%====================================================================================================
%====================================================================================================

By this, I mean the BO ansatz for the wavefunction, 
\be
\Psi = \psi(h_{\alpha\beta})
|\zeta_{\sfj}(h_{\alpha\beta}, \psi_{\Gamma})\rangle
\mbox{ } ,
\label{BO}
\ee  
and the whole package of approximations conventionally made alongside it, 
only one of which is the direct analogue of {\sl the} BO approximation.

The inner product used below is $\langle\zeta|\zeta^{\prime}\rangle = 
\int \d\nL_{\Delta} \zeta^*(h_{\alpha\beta}, \psi_{\Gamma})\zeta^{\prime}(h_{\alpha\beta}, \psi_{\Gamma})$.   
In this paper, the covariant derivative is 
\be
{{\cal D}_{\sh}}^{\alpha\beta} = {\delta_{\sh}}^{\alpha\beta} + i{{\cal A}_{\sh}}^{\alpha\beta} \mbox{ } ,
\label{Bercov}
\ee
with conjugate 
\be
{{\cal D}_{\sh}^*}^{\alpha\beta} = {\delta_{\sh}}^{\alpha\beta} - i{{\cal A}_{\sh}}^{\alpha\beta} \mbox{ } .
\label{covstar}
\ee  
The connection therein is the counterpart of Berry's connection that the MCMSF space induces on Riem,  
\be
{\cal A}^{\alpha\beta} = 
-i\left< \zeta_{\sfj}  \right| {\delta_{\sh}}^{\alpha\beta} \left|\zeta_{\sfj} \right> = 
- i\int \d \psi \zeta_{\sn}^*(h_{\alpha\beta}, \psi){\delta_{\sh}}^{\alpha\beta}
\zeta_{\sn}(h_{\alpha\beta}, \psi) 
\mbox{ } .
\ee
All of this assumes a nondegenerate quantum cosmological state.\fn{If this were not the case, 
one would have to consider the geometrodynamical counterpart of footnote I.8.}
%crossfootnote.   

The following correspondence permits uplift of the identities (I.46--49) to this paper:
\be 
\pa_{\sH} \longrightarrow \delta_{\sh}            \mbox{ } , \mbox{ } 
A_{\sH}   \longrightarrow {\cal A}_{\sh}          \mbox{ } , \mbox{ }  
D_{\sH}   \longrightarrow {\cal D}_{\sh}          \mbox{ } , \mbox{ } 
\mbox{Paper I's notion of }\mbox{}_{\sM}  \longrightarrow \mbox{Paper II's notion of }\mbox{}_{\sM}  \mbox{ } . \mbox{ } 
\label{corr}
\ee  
N.B. using these equations is crucially subject to the 
configuration space metric depends on the H d.o.f.'s alone, else one is offset by ordering problems.

The first few equations of this Paper's formalism for geometrodynamics are: 
$$
\stackrel{\underline{\mbox{preliminary}}}{\underline{\mbox{equations}}} 
\hspace{0.05in}
\stackrel{\mbox{$-\hbar^2\mbox{}_{\sM^{-1}}||\delta_{\sh}||^2\lfloor | \zeta \rangle\psi\rfloor$}  }
         {+\hat{h}|\zeta\rangle\psi = \sqrt{h}\tilde{\Lambda}|\zeta\rangle\psi} 
\hspace{0.05in}
{\longrightarrow}
\hspace{0.05in}
\stackrel{          \mbox{    $-\hbar^2
\left\{
|\zeta\rangle \mbox{}_{\sM^{-1}}   ||\delta_{\sh}||^2    \psi + 
2_{\sM^{-1}}(\lfloor\delta_{\sh}\psi\rfloor , \delta_{\sh}| \zeta\rangle) +
\right.  $    }        } 
{           
\left. 
\psi_{\sM^{-1}}  ||\delta_{\sM^{-1}}||^2|\zeta\rangle                             
\right\}                                                                                        
+ \hat{h}|\zeta\rangle\psi = \sqrt{h}\tilde{\Lambda}|\zeta\rangle \psi           }                                                                                                                     
\hspace{0.05in}
{\longrightarrow}
\hspace{0.05in}
\stackrel{\mbox{$-\hbar^2|\zeta\rangle\mbox{}_{\sM^{-1}}||\delta_{\sh}||^2\psi  + \hat{h}|\zeta\rangle$} }
         {  + 2_{\sM^{-1}}(\lfloor\delta_{\sh}\psi\rfloor , \delta_{\sh}| \zeta\rangle) 
          = \sqrt{h}\tilde{\Lambda}|\zeta\rangle\psi}
$$
$$
\hspace{2in} \downarrow \hspace{3in} \hspace{1.5in} \downarrow \hspace{1.5in} 
$$
$$
\underline{\mbox{H-equations}}
\stackrel{\mbox{$-\hbar^2\langle\zeta|_{\sM^{-1}}||\delta_{\sh}||^2 \lfloor \psi |\zeta \rangle \rfloor$}}
         {+ {\cal O}\psi = \sqrt{h}\tilde{\Lambda}\psi                                                }
\hspace{2.5in}
\stackrel{\mbox{(Banks}}
         {\mbox{equation \cite{Banks})}}
\stackrel{\mbox{$-\hbar^2|\zeta\rangle_{\sM^{-1}}||\delta_{\sh}||^2\psi + {\cal O}|\zeta\rangle \psi$}}
{+ 2_{\sM^{-1}}(\lfloor\delta_{\sh}\psi\rfloor , \delta_{\sh}| \zeta\rangle) = 
\sqrt{h}\tilde{\Lambda}|\zeta\rangle \psi} 
$$
$$
\hspace{2in} \downarrow \hspace{1.5in} \hspace{1.5in} \hspace{1.5in} \uparrow \hspace{1.5in} 
$$
\be
- \hbar^2\mbox{}_{\sM^{-1}}||{\cal D}_{\sh}||^2\psi + {\cal E} + {\cal O}\}\psi = \sqrt{h}\tilde{\Lambda}\psi 
\stackrel{\mbox{(H-equation in}}
         {\stackrel{\mbox{Berry form}}
                   {\mbox{generalized to Riem)}}            }
\longrightarrow
\stackrel{        \mbox{(H-equation in Berry--Simon}        }
         {        \stackrel{            \mbox{geometrical form}            }
                           {            \mbox{generalized to Riem)}             }        }
 - \hbar^2\mbox{}_{\sM^{-1}}||{\cal D}_{\sh}||^2\psi + {\cal O}\psi = \sqrt{h}\tilde{\Lambda}\psi \mbox{ } .
\label{boarray}
\ee
Here, ${\cal E}$ is the generalization on Riem of the electric term, which is the M-trace 
of $<Q^{\alpha\beta\Gp\Dp}>$, which is the generalization on Riem of Berry's quantum geometric tensor 
\cite{Berry89}
\be
<Q^{\alpha\beta\Gp\Dp}> = 
\mbox{Re}\{\langle \lfloor{\delta_{\sh}}^{\alpha \Gp}\zeta\rfloor| \{ 1 - P_{\zeta} \}
\lfloor {\delta_{\sh}}^{\beta\Dp} | \zeta \rfloor \rangle \}\psi \mbox{ } , 
\ee
and $P_{\zeta}$ is the projector $|\zeta\rangle\langle\zeta|$.  
${\cal O}$ is the `$h_{\alpha\beta}$-parameter dependent eigenvalue' of 
$\hat{h} \equiv {\cal H}_{\sh\phi} + {\cal V}_{\phi}$: 
\be
\hat{h}(h_{\alpha\beta}, \phi_{\Gamma}, \np^{\Gamma}) | \psi(h_{\alpha\beta}, \phi_{\Gamma}) \rangle = 
{\cal O}(h_{\alpha\beta}) | \psi(h_{\alpha\beta}, \phi_{\Gamma}) \rangle \mbox{ }  .
\ee
This is only a consistent procedure if the off-diagonal components of the matrix 
${\cal O}_{\mbox{\scriptsize jl}} = \langle\zeta_{\sj} | \hat{h} | \zeta_{\sll}\rangle$ are negligible: 
\be
\mbox{for } \nj \neq \nl \mbox{ , } 
\left|
{\cal O}_{\mbox{\scriptsize jl}}/{\cal O}_{\mbox{\scriptsize jj}}
\right|
= \varepsilon_{\sBO} \mbox{ , small (BO approximation on Riem $ \times $ matter configuration space) }.
\ee
%*introduce a symbol for that, is the blockwise expression.  

The diagram covers all of: the Banks analogue of BO's scheme modified by cross-term keeping: 
ABC,\fn{This path follows the Born--Fock neglect of what are now known to be connection terms.
The argument that it is not necessary to keep these in 
minisuperspace quantum cosmology with 1 H d.o.f. so that loops aren't possible in the 
H-subconfiguration space is overriden by the considerations D4 that relative phase connection terms are 
appropriate, as relative phase does not require a loop to build up over.} 
the Riem analogue of Berry's scheme DE (of which e.g. Brout--Venturi's scheme \cite{BV89} 
is the minisuperspace version), and the recovery of Banks's scheme from this, FG.  
The merits of scheme DE as opposed to Banks' scheme carry over from the corresponding 
discussion of Berry's and BO's schemes on p. 9 of I.

Banks's scheme ABC involves, respectively: 
expanding by (I.46) under correspondence (\ref{corr}),  
two adiabatic neglects $\varepsilon_{\sa \sw 3}$, $\varepsilon_{\sa \sw 7\scr}$ small, 
defining ${\cal O}$, premultiplication by $\langle \zeta |$, 
the acceptability of which is underlied by a 
diagonal dominance condition 
(over the likewise-defined  ${\cal O}_{\sj\sll}$ now built with distinct $\langle\zeta_{\sj}|$, $|\zeta_{\sll}\rangle$):   
\be
\mbox{for } \nn \neq \nl \mbox{ } , 
\left|{\cal O}_{\mbox{\scriptsize jl}}/{\cal O}_{\mbox{\scriptsize jj}}  
\right|
= \epsilon_{\sBO} \mbox{ , small (BO approximation) },
\ee
and finally making use of the normalization of $|\zeta\rangle$.
Berry's move E is via identity (I.50) under correspondence (\ref{corr}),   
and amounts to casting the H-equation in a geometrical form.
The context for this is an adiabatic loop in phase space, 
whence this scheme is underlied by being in a classically-adiabatic regime
(i.e. that classical H-processes are much slower than classical L-processes),\fn{
As most of the potentially small quantities have direct analogues in Paper I, I refer the reader 
to Sec I.2--6 for the meaning of the majority of the present paper's suffix notation for 
small quantities.}   
\be
{\Omega_{\sh}}/{\omega_{\phi}} = \varepsilon_{\sa} \mbox{ } 
\ee
for $\Omega_{\sh}$ and $\omega_{\phi}$ `characteristic frequencies' of the gravitational and matter 
subsystems respectively.  
This amounts to a comparison of inverse lengths, requiring 
$1/\sqrt{R}$ (curvature scale) and $1/\sqrt{\Lambda}$ (cosmological constant scale) 
to be much larger than the scalar field inhomogeneity $|\pa\phi|$ and Compton wavelength.  
This corresponds to a (unusual) classicality condition at late times (see \cite{InflationBG} for 
a related discussion).  
Moves D and E can be encapsulated together as another `diagonal dominance',
\be
\mbox{ for } \nn \neq \nl \mbox{ } , \mbox{ } 
\varepsilon_{\sd\sBOB} = \left| \{{\cal O}_{\sll\sj} + {\cal E}_{\sll\sj}\}/ 
\{{\cal O}_{\sj\sj} + {\cal E}_{\sj\sj}\}
\right| 
\mbox{ , small .} 
\ee

Move F is via considering the quantum correction potential ${\cal E}$ 
to be dominated by usually ${\cal O}$ but just as well by 
$\mbox{}_{\sM^{-1}}||\delta_{\sh}||^2|\zeta\rangle$: neglecting 
$\varepsilon_{\sa\sw 1}$ and $\varepsilon_{\sa\sw 2}$.  
Then recovering the BO equation by move G involves neglecting 
$\varepsilon_{\sa\sw 1}$, $\varepsilon_{\sa \sw 2}$ compensatorily (such that the whole of FG does not require 
these two approximations to be made) and also neglecting $\varepsilon_{\sa \sw 4}$ and $\varepsilon_{\sa \sw 8}$.  
The last 2 of these are in close correspondence with the terms neglected in move B.
However, arriving at Banks equation via the long path again requires more work, 
reflecting that making BO's adiabatic assumptions and forming H-equations 
are non-commuting procedures.

%========================================================================================================
\noindent{\underline{L-equations}}
%========================================================================================================
%
%
%
One considers next equations of the form 
$\{ \mbox{Preliminary equation}\} - \{\mbox{H-equation}\}|\zeta\rangle$, which are prima facie 
fluctuation equations.  
From the top LHS version of preliminary equation in (\ref{boarray}) and the 
Berry version of the H-equation, this takes the form 
\be
\frac{1}{\hbar^2}\overline{\hat{h}} = \mbox{ }_{\sM^{-1}}||\delta_{\sh}||^2\lfloor \psi|\zeta\rangle \rfloor 
-  \left\{ \mbox{ }_{\sM^{-1}}||{\cal D}{\sh}||^2\lfloor\psi\rfloor 
+  \langle _{\sM^{-1}}||D^{*}_{\sh}||^2\lfloor\psi\rfloor\rangle 
\right\}
|\zeta\rangle \mbox{ } .  
\label{FLE}
\ee
Alternatively, rearranging by (I.53) under correspondence (34), this takes the form 
\be
\left\{\overline{\frac{1}{\hbar^2}\hat{h} - \mbox{ }_{\sM^{-1}}||{\cal D}^*_{\sh}||^2 } \right\}|\zeta\rangle = 
\frac{2}{\psi}\mbox{}_{\sM^{-1}}(\lfloor {\cal D}_{\sh}\psi\rfloor, {\cal D}_{\sh}^*|\zeta\rangle) 
\hspace{0.4in} 
\stackrel{\mbox{(generalization of  Brout--Venturi's \cite{BV89}}}{\mbox{minisuperspace fluctuation equation)}} 
\mbox{ } .
\label{TLP}
\ee

The quantum cosmological point about the cross-term-possessing L-equations however 
\cite{LR79, Banks, HallHaw}, is that, as covered in Sec 7, 
such a fluctuation equation can be rearranged to form a time-dependent wave equation (TDWE) for the L-subsystem, 
with respect to an (approximate) time induced by the H-subsystem.  
The kind of cross-term on the RHS is crucial for this quantum cosmological scheme; note that this is 
entirely thrown away in the standard BO approach to QM.

The expanded L-equation  
$$
\frac{1}{\hbar^2}
\left\{
\stackrel{\mbox{$- \hbar^2\mbox{ }_{\sm^{-1}}||\delta_{\phi}||^2$}}
{+ \hbar^2 \langle\mbox{$ \mbox{}_{\sm^{-1}}||\delta_{\phi}||^2$}\rangle}
\stackrel{\mbox{$- \tilde{{\cal R}}$}} 
{+\langle\mbox{$ \tilde{\cal R}$}\rangle}
\stackrel{\mbox{$+ {\cal V}_{\phi}$}}
{-\langle\mbox{$ {\cal V}_{\phi}$}\rangle} 
\stackrel{\mbox{$+ {\cal I}_{\sh\phi}$}}
{-\langle\mbox{$ {\cal I}_{\sh\phi}$}\rangle}
\right\} 
|\zeta\rangle - 
\left\{
\stackrel{\mbox{$\mbox{ }_{\sM^{-1}}||\delta_{\sh}||^2$}} 
{-\langle\mbox{$\mbox{ }_{\sM^{-1}}||\delta_{\sh}||^2$}\rangle} 
\stackrel{\mbox{$- 2i_{\sM^{-1}}({\cal A}_{\sh}, \delta_{\sh})$}}
{+ 2i\langle\mbox{$ \mbox{}_{\sM^{-1}}({\cal A}_{\sh}, \delta_{\sh})$}\rangle}
\stackrel{\mbox{$- i_{\sM^{-1}}(\lfloor\delta_{\sh}, {\cal A}{\sh}\rfloor)$}} 
{+\langle\mbox{$i \mbox{}_{\sM^{-1}}(\lfloor\delta_{\sh}, {\cal A}{\sh}\rfloor)$}\rangle} 
\stackrel{\mbox{$- \mbox{}_{\sM^{-1}}||{\cal A}{\sh}||^2$}}
{+\langle\mbox{$ \mbox{}_{\sM^{-1}}||{\cal A}{\sh}||^2$}\rangle}
\right\}
|\zeta\rangle
$$
\be
= \frac{2}{\psi}
\left\{
\mbox{ }_{\sM^{-1}}(\lfloor \delta_{\sh}\psi\rfloor, \delta_{\sh}|\zeta\rangle) 
+ \mbox{ }_{\sM^{-1}}||{\cal A}{\sh}||^2\psi|\zeta\rangle 
+ i_{\sM^{-1}}({\cal A}{\sh}, \psi\delta_{\sh}|\zeta\rangle - |\zeta\rangle \delta_{\sh}\psi)
\right\}
\mbox{ }
\ee
is useful for the below discussion of approximations.
Again, the second, sixth and seventh columns of the LHS 
cancel out because the weightings of $\langle \mbox{ } | \mbox{ } | \mbox{ } \rangle$ 
are functionals of the metric alone and hence can be pulled outside the functional integrals 
over $\psi$.    
[On the other hand, the corresponding expanding out of the generalized Berry H-equation 
merely involves applying (I.47) under the correspondence (\ref{corr}) 
to it, so I do not provide it.]

%========================================================================================================
\noindent\underline{Corresponding momentum constraint equations.} 
%========================================================================================================
%
%
%
The above scheme serves for minisuperspace models, but beyond these one has to handle also the momentum 
constraint.
One natural approach to the momentum constraint is to treat it in parallel with how the 
Hamiltonian constraint is treated.\fn{Another 
%%%%%%%%%%%%%%%%%%%%%%%%%%%%%%%%%%%%%%%%%%%%%%%%%%%%%%%%%%%%%%%%%%%%%%%%%%%%%%%%%%%%%%%%%%%%%%%%%%%%%%%%%
way would be to try to solve these classically, substitute them into 
the quadratic constraint at the classical level and then only have a quadratic constraint to deal with 
-- a superspace quantization type scheme \cite{I76}
 with a 1- (rather than 4-)component many-fingered time.}
%%%%%%%%%%%%%%%%%%%%%%%%%%%%%%%%%%%%%%%%%%%%%%%%%%%%%%%%%%%%%%%%%%%%%%%%%%%%%%%%%%%%%%%%%%%%%%%%%%%%%%%%%
The analogue of the above `cycle' is then:

\noindent{\mbox{\underline{preliminary momentum constraint equations}}}
$$
\stackrel{\mbox{$\left\{-2\frac{\hbar}{i}\nabla_{\alpha}{{\delta}_{\sh}}\mbox{}^{\alpha\beta}\right.$}} 
         {\mbox{$\left. + \frac{\hbar}{i}\lfloor \pa^{\beta}\phi_{\Gamma} \rfloor {{\delta}_{\phi}}^{\Gamma}
\right\}\psi|\zeta\rangle = 0$}}
\hspace{0.05in}
\stackrel{\mbox{Step A}}{\longrightarrow}
\hspace{0.05in}
\stackrel{\mbox{$-2\frac{\hbar}{i}
           \left\{
          \psi\nabla_{\alpha}{{\delta}_{\sh}}\mbox{}^{\alpha\beta}|\zeta\rangle
        + \lfloor \nabla_{\alpha}\psi \rfloor {{\delta}_{\sh}}\mbox{}^{\alpha\beta}|\zeta\rangle
        + \lfloor \nabla_{\alpha}|\zeta\rangle \rfloor {{\delta}_{\sh}}\mbox{}^{\alpha\beta} \psi \right.$ +}}
        {\mbox{$\left.|\zeta\rangle \nabla_{\alpha} {{\delta}_{\sh}}\mbox{}^{\alpha\beta}\psi
          \right\}
        + \frac{\hbar}{i}\lfloor \pa^{\beta}\phi_{\Gamma} \rfloor \psi {\delta}_{\phi}\mbox{}^{\Gamma}|\zeta\rangle = 0$}}  
\hspace{0.05in}
\stackrel{\mbox{Step B}}{\longrightarrow}
\hspace{0.05in}
\stackrel{\mbox{$-2\frac{\hbar}{i}
          \psi\nabla_{\alpha}{{\delta}_{\sh}}\mbox{}^{\alpha\beta}|\zeta\rangle$}}
        {\mbox{$ + \frac{\hbar}{i}\lfloor \pa^{\beta}\phi_{\Gamma} \rfloor \psi {\delta}_{\phi}\mbox{}^{\Gamma}|\zeta\rangle = 0$}} 
$$
$$
\hspace{1in} \mbox{Step D} \downarrow \hspace{2in} \hspace{1.5in} \hspace{1in} \mbox{Step C } \downarrow \hspace{1.5in} 
$$
$$
\noindent{\mbox{\underline{momentum constraint H-equations}}} \hspace{6in}
$$
\be
\frac{\hbar}{i}\left\{-2\langle\zeta|\nabla_{\alpha}{{\delta}_{\sh}}\mbox{}^{\alpha\beta}\psi|\zeta\rangle + 
\langle\zeta|\lfloor \pa^{\beta}\phi_{\Gamma} \rfloor {\delta}_{\phi}\mbox{}^{\Gamma}
\psi|\zeta\rangle
\right\} = 0
\hspace{1.05in}
\stackrel{\mbox{Step F}}{\longrightarrow}
\hspace{1.05in}
\stackrel{\mbox{$-2\frac{\hbar}{i}
          \langle\zeta|\psi\nabla_{\alpha}{{\delta}_{\sh}}\mbox{}^{\alpha\beta}|\zeta\rangle$}}
        {\mbox{$ + \frac{\hbar}{i}\langle\zeta|\lfloor \pa^{\beta}\phi_{\Gamma} \rfloor \psi {\delta}_{\phi}\mbox{}^{\Gamma}|\zeta\rangle = 0$}} 
\mbox{ } .
\ee
Note that the last term is $\langle {{\cal M}^{\beta}}_{\phi}\rangle\psi$.  

By passage from the preliminary equations (line 1) to the H-equations, what is meant is 

\noindent 1) define 
\be
{\cal M}_{\sfj\sfl} = \langle\zeta_{\sfj}|\hat{\cal M}|\zeta_{\sfl}\rangle \mbox{ } .
\ee 
2) Premultiply the preliminary equation by 
$\langle \zeta |$, the acceptability of which is underlied by the diagonal dominance   
\be
\mbox{for } \fj \neq \fl \mbox{ } , 
\left|{\cal M}_{\sfj\sfl}/{\cal M}_{\sfj\sfj}  
\right|
= \varepsilon_{\sd(\smom)} \mbox{ , small }.  
\ee
3) Make use of the normalization of $|\zeta\rangle$.
Also, both paths involve 3 adiabatic neglects analogous in pairs between the 2 paths, 
so this diagram commutes in the same sense that its ZAM counterpart does.  
While some kinds of cross-term arise, these are not of the same kind as crucial chroniferous one 
obtained from the quadratic constraint, similarly to what occurs with the ZAM constraint, the 
momentum constraint does not give rise to another TDWE but rather provides a piece to the quadratic 
constraint TDWE.    
Where the treatment of the momentum constraint does present a problem absent from the treatment 
of the simpler ZAM constraint is that the constraint's differential operator obstructs the 
$\delta_{\sh}\psi + \psi\langle\zeta|\delta_{\sh}|\zeta\rangle = {\cal D}_{\sh}\psi$ grouping, 
preventing one from attaining at least straightforwardly a Berry-like geometrization away from minisuperspace.  
Thus Berry's manifestly geometrical scheme requires a subtle generalization 
if it is to apply to general contrained field theories.

%========================================================================================================
\noindent{\underline{Momentum constraint fluctuation L-equation.}}
%========================================================================================================
%
%
%
\be
\overline{    {\cal M}_{\sh}\mbox{}^{\beta}    }|\zeta\rangle + 
\overline{    {\cal M}_{\phi}\mbox{}^{\beta}   }|\zeta\rangle = 0
\label{MCL}
\ee
for
\be
{\cal M}_{\sh}  \mbox{}^{\beta} = \frac{\hbar}{i}
\left\{
\psi \overline{    \nabla_{\alpha}  {{\delta}_{\sh}}^{\alpha\beta}    } + 
\lfloor\nabla_{\alpha}\psi\rfloor \overline{    {{\delta}_{\sh}}^{\alpha\beta}    } + 
\overline{    {{\delta}_{\sh}}^{\alpha\beta}\psi \lfloor\nabla_{\alpha}\rfloor    }
\right\} 
\mbox{ } .  
\ee
%As $\psi$ is of H alone, can take inside the overline.
%
This is easy to strip down, by (\ref{covstar}).  
Finally, note that the BO type working's L-equation reads simply 
\be
\overline{{\cal M}_{\phi}\mbox{}^{\beta}}|\zeta\rangle = 0 \mbox{ } ,  
\ee
which amounts to a lack of geometry--matter interaction.

%========================================================================================================
\noindent\underline{Approximations} 
%========================================================================================================
%
%
%
I next list and characterize the subsequent plethora of approximations, many of which are 
made in the semiclassical quantum cosmology literature.  
One should interconvert by $\phi \longrightarrow \psi$, 
$\chi \longrightarrow \zeta$, $\fF \longrightarrow {\cal F}$ and (\ref{corr}). 
Consult the previous paper for a more of the notation and significance  
for those potentially small quantities which have have direct analogues there, 
while, I now use `mom' for terms orginating in the momentum constraint.

I maintain the policy of intending to build from primitives, 
albeit this cannot be completed yet in this Section.  
Fairly primitive quantities that occur in the various equations and might be considered to be 
small/negligible are
$$
\left| {\cal A}_{\sh}^2/ \delta_{\phi}\mbox{}^2|\zeta\rangle \right| = \varepsilon_{\sa \sp 1}
\mbox{ } , \mbox{ } 
\left| {\delta_{\sh}}^2|\zeta\rangle / \delta_{\phi}\mbox{}^2|\zeta\rangle \right| = \varepsilon_{\sa \sp 3}
\mbox{ } , \mbox{ }
|{\langle\delta_{\sh}\mbox{}^2\rangle|\zeta\rangle}/\delta_{\phi}\mbox{}^2|\zeta\rangle| = \varepsilon_{\sa\sp 4} 
\mbox{ } , \mbox{ }
$$
$$
\left|
{\cal A}_{\sh}^2/\langle \delta_{\phi}\mbox{}^2\rangle
\right|
= \varepsilon_{\sa \sp^{\prime}1}
\mbox{ } , \mbox{ }
\left|
\mbox{}_{\sm^{-1}}|| |\delta_{\sh}\zeta\rangle ||^2/\langle \delta_{\phi}\mbox{}^2\rangle 
\right| 
= \varepsilon_{\sa \sp^{\prime}2}
\mbox{ } , \mbox{ }
|{{\cal A}_{\sh}\delta_{\sh}|\zeta\rangle}/\delta_{\phi}\mbox{}^2|\zeta\rangle| = 
\varepsilon_{\sa\sp 5} 
\mbox{ } , \mbox{ }
|\langle {\cal A}_{\sh} \delta_{\sh}\rangle| \zeta\rangle/{\delta_{\phi}}^2|\zeta\rangle| = \varepsilon_{\sa\sp 6}  
\mbox{ } , \mbox{ }
$$
\be
\left|
\lfloor \delta_{\sh}\psi \rfloor \delta_{\sh} |\zeta\rangle/\delta_{\phi}\mbox{}^2|\zeta\rangle
\right| 
= \varepsilon_{\sa \sm 7\scr}
\mbox{ } , \mbox{ }
\left|
{\cal A}_{\sh}\delta_{\sh}|\zeta\rangle/|\delta_{\phi}\mbox{}^2\zeta\rangle
\right| =
\left|
\langle \zeta | \lfloor \delta_{\sh} |\zeta \rfloor\rangle \delta_{\sh}\psi/\delta_{\phi}\mbox{}^2|\zeta\rangle
\right| 
= \varepsilon_{\sa \sm 8\scr}
\mbox{ } , \mbox{ }
 \ee
\be
\left|
\langle {\cal V}_{\phi} \rangle/\hbar^2 \delta_{\phi}\mbox{}^2 |\zeta \rangle 
\right| 
= \varepsilon_{\sL 1}
\mbox{ } , \mbox{ }
\left|
\langle {\cal I}_{\sh\phi} \rangle/\hbar^2 \delta_{\phi}\mbox{}^2 |\zeta \rangle 
\right| 
= \varepsilon_{\sL 2}
\mbox{ } , \mbox{ }
\left|
\langle \delta^2_{\phi} \rangle/\hbar^2 \delta_{\phi}\mbox{}^2 |\zeta \rangle 
\right| 
= \varepsilon_{\sL 3}
\mbox{ } . \mbox{ }
\ee
Also, there are 9 quantities arising from the momentum constaint: 
3 along the top path, with $\langle\zeta|$ times numerator 
versions of each of these along the bottom path 
and 3 averaged terms with no lead term to compare to from the L-equation.  
These quantities have some features of the proto-WKB type, but are additionally intermixed with 
conditions involving spatial derivatives of the wavefunction ansatz's pieces.
$$
| \mbox{ } || {\cal A}_{\sh} ||^2/\delta_{\sh}\mbox{}^2\psi | = \varepsilon_{\sa \sw 1}
\mbox{ , }
| _{\sm^{-1}}|| \delta_{\sh}\zeta ||^2/\delta_{\sh}\mbox{}^2\psi| = \varepsilon_{\sa \sw 2} 
\mbox{ , }
|\delta_{\sh}\mbox{}^2|\zeta\rangle / \delta_{\sh}\mbox{}^2\psi| = \varepsilon_{\sa \sw 3}
\mbox{ , }
|\lfloor \delta_{\sh}\psi\rfloor \delta_{\sh}|\zeta\rangle/\delta_{\sh}\mbox{}^2\psi| = \varepsilon_{\sa \sw 4} 
\mbox{ } ,
$$
\be
\left|
\lfloor\delta_{\sh}\psi\rfloor\delta_{\sh}|\zeta\rangle/\delta_{\sh}\mbox{}^2|\zeta\rangle
\right| 
= \varepsilon_{\sa \sw 7\scr} 
\mbox{ } , \mbox{ }
\left|
{\cal A}_{\sh}\delta_{\sh}|\zeta\rangle/\delta_{\sh}\mbox{}^2|\zeta\rangle
\right| 
=  
\left|
\langle\zeta|\lfloor\delta_{\sh}|\zeta\rfloor\rangle\delta_{\sh}\psi/\delta_{\sh}\mbox{}^2|\zeta\rangle
\right|
= 
\varepsilon_{\sa \sw 8\scr} 
\mbox{ } .
\ee

As regards what happened to the abovementioned diagonal terms, 
$\varepsilon_{\sd\sBO}$ and $\varepsilon_{\sd\sAM}$
are not per se adiabatic and are kept as primitives. 
To relate to the previously mentioned quantity $\varepsilon_{\sd\sBOB}$ to primitives in use above,
I expand its definition (assuming the BO term $\varepsilon_{\sBO}$ is largest therein):
$\varepsilon_{\sd\sBOB} = \{\varepsilon_{\sd\sBO} + \varepsilon_{\sd\sBOB^{\prime\prime}}\}
                     \{  1 + O\{ \varepsilon_{\sd\sBOB^{\prime}}\}$
for $\varepsilon_{\sd\sBOB^{\prime}} = {\cal E}_{\sj\sj}/{\cal O}_{\sj\sj}$ 
and $\varepsilon_{\sd\sBOB^{\prime\prime}} = {\cal E}_{\sll\sj}/{\cal O}_{\sll\sj}$.  
It is from expanding these out from the definitions of ${\cal E}$ and ${\cal O}$ that 
$\varepsilon_{\sp 1^{\prime}}$, $\varepsilon_{\sp 2^{\prime}}$, 
$\varepsilon_{\sp^{\prime}2^{\prime}}$ and $\varepsilon_{\sp^{\prime}2^{\prime}}$ arise, alongside mass 
factors.
Overall, 
$\varepsilon_{\sd\sBOB} = 
\varepsilon_{\sd\sBO} + \varepsilon_{\sH\sL} 
\{ \varepsilon_{\sa \sp^{\prime}1^{\prime}} + \varepsilon_{\sa\sp^{\prime}1^{\prime}} \} +
\varepsilon_{\sd\sBO} \varepsilon_{\sH\sL} \{ \varepsilon_{\sa\sp^{\prime}1} + 
\varepsilon_{\sa\sp^{\prime}2} \} +
\varepsilon_{\sH\sL} \{ \varepsilon_{\sa\sp^{\prime}1^{\prime}} + \varepsilon_{\sa\sp^{\prime}1^{\prime}} \}
O(\varepsilon_{\Delta \sM}, \varepsilon_{\Delta \sm}) + ...$,
though exactly what is kept in the expansions depends on the relative sizes of the 
various `small $\varepsilon$ quantities'

Finally, the table of properties in Paper I passes over to this Paper by swapping the ZAM entry 
for the corresponding mom entry.  Each $\varepsilon$ that has a corresponding $\epsilon$ at the end 
of Sec I.5 shares its properties as listed there.

%====================================================================================================
%====================================================================================================
\section{The WKB procedure}
%====================================================================================================
%====================================================================================================

I take this to consist of the subsequent H-wavefunction ansatz 
$$
\psi = \mbox{exp}
\left(
iM^2_{\mbox{\scriptsize Pl}}{\cal F}
\left(
h_{\alpha\beta}
\right)
/\hbar
\right)
$$
and some habitually-associated approximations.

Then the bottom right-hand corner H-equation in (\ref{boarray}) expanded by (I.47) 
under correspondence (\ref{corr}) becomes 
\be
\frac{\hbar}{c}
\left\{ 
M_{\mbox{\scriptsize Pl}}^2 \mbox{}_{\sM^{-1}}||\delta_{\sh}{\cal F}||^2 - 
i\hbar \mbox{}_{\sM^{-1}}||\delta_{\sh}||^2{\cal F} + 
2\hbar\mbox{}_{{\cal G}}({\cal A}{\sh}, \delta_{\sh}{\cal F})
- \frac{\hbar^2}{M_{\Pl}^2}\mbox{}_{{\cal G}}(\delta_{\sh}, {\cal A}{\sh}) + 
\frac{\hbar^2}{M_{\Pl}^2}\mbox{}_{{\cal G}}||{\cal A}{\sh}||^2 
\right\}
+ {\cal E}(M_{\Pl}) + {\cal O}(M_{\Pl}) = \frac{c^3 M_{\Pl}^2}{\hbar}\sqrt{h}{\Lambda} 
\mbox{ } ,
\ee
\mbox{ } \mbox{ } Moreover, the L-equation (\ref{TLP}) becomes

\noindent
\be
\left\{
\frac{1}{\hbar^2}\overline{\hat{h}}(M_{\Pl}) + \hat{R}(M_{\Pl})
\right\}
|\zeta\rangle = \frac{2 iM_{\mbox{\scriptsize Pl}}^2}{c \hbar} 
\left\{
\mbox{}_{{\cal G}}(\lfloor\delta_{\sh}{\cal F}\rfloor,\delta_{\sh})|\zeta\rangle - 
i_{{\cal G}}(\lfloor\delta_{\sh}{\cal F}\rfloor, {\cal A}{\sh})|\zeta\rangle 
\right\} 
\mbox{ } , 
\label{preTDSE}
\ee
which is arranged so that the RHS exclusively and exhaustively isolates the cross-terms, 
all other types of correction terms being bundled into the LHS's `remainder operator' 
\be
\hat{R} = \frac{\hbar}{cM_{\Pl}^2}
\left\{
i_{{\cal G}}({\cal A}{\sh}, \delta_{\sh}) - 2i\langle_{{\cal G}}
({\cal A}{\sh}, \delta_{\sh})\rangle - 
\mbox{}_{{\cal G}}||{\cal A}{\sh}||^2 - \overline{\mbox{}_{{\cal G}}||\delta_{\sh}||^2} 
\right\} 
\mbox{ } .  
\ee

%=======================================================================================================
\noindent\underline{Corresponding  WKB momentum constraint equations.}  
%=======================================================================================================
%
%
%
In the case of spatially-nontrivial geometrodyamics, 
there is also a momentum H-equation, 
$$
-2
\left\{
\langle \zeta|\nabla_{\alpha}{{\delta}_{\sh}}^{\alpha\beta} |\zeta \rangle + 
\frac{iM_{\Pl}^2}{\hbar}\lfloor\nabla_{\alpha}{\cal F}\rfloor \langle \zeta| {{\delta}_{\sh}}^{\alpha\beta}
|\zeta \rangle + 
\frac{iM_{\Pl}^2}{\hbar}\langle \zeta| \lfloor \nabla_{\alpha} |\zeta \rfloor\rangle 
{{\delta}_{\sh}}^{\alpha\beta}{\cal F} + 
\frac{iM_{\Pl}^2}{\hbar}\nabla_{\alpha}{\delta_{\sh}}^{\alpha\beta}{\cal F} - 
\frac{M_{\Pl}^4}{\hbar^2}\lfloor \nabla_{\alpha}{\cal F} \rfloor {{\delta}_{\sh}}^{\alpha\beta}{\cal F}
\right\}
$$
\be
+ \langle \zeta|\lfloor\pa^{\beta}\phi_{\Gamma}\rfloor{\delta_{\phi}}^{\Gamma}|\zeta \rangle = 0 
\mbox{ } .  
\ee
Moreover, the momentum L-equation has its previous form (\ref{MCL}), but now with
\be
\overline{    {\cal M}    }_{\sh}\mbox{}^{\beta} =
-2\frac{\hbar}{i}
\left\{
\overline{    \nabla_{\alpha}  {{\delta}}_{\sh}\mbox{}^{\alpha\beta}    } + 
\frac{    iM_{\Pl}    }{    \hbar    }
\lfloor\nabla_{\alpha}{\cal F}\rfloor\overline{    {\delta}_{\sh}\mbox{}^{\alpha\beta}    }|\zeta\rangle  + 
\frac{    iM_{\Pl}    }{    \hbar    }
\lfloor\overline{    {\delta}_{\sh}\mbox{}^{\alpha\beta}    {\cal F}\rfloor\nabla_{\alpha}    }
\right\}
\mbox{ } .
\ee

%=======================================================================================================
\noindent\underline{Approximations.} 
%=======================================================================================================
%
%
%
Upon adopting the WKB ansatz, the p and L criteria remain as in Sec 5, 
while that ${\pounds}$ and w criteria are modified by the adoption of the WKB ansatz.  
One now has ${\cal F}$-change rather than $\psi$-change with respect to H, alongside some power of 
$M_{\mbox{\scriptsize Pl}}/\hbar$ which ensure that one continues to talk about dimensionless ratios.  
Again, I call the resulting quantities `g' and `W'.
There continues to be a relation between these various quantities:  
$\varepsilon_{\sg}$/$\varepsilon_{\sw}$ = $\varepsilon_{\sp}$.   
The `WKB approximation' is that the ${\cal F}$ is slowly varying with respect to $h_{\alpha\beta}$.  
This amounts to the following string of approximations. 
Firstly, there is the typical WKB assumption that 
\be
\left|\frac{\hbar\delta_{\sh}\mbox{}^2{\cal F}}{M_{\mbox{\scriptsize Pl}}^2 |\delta_{\sh}{\cal F}|^2}\right| = \varepsilon_{\ssWKB} \mbox{ , small , }
\ee
which is an approximation type lying outside the p, W, g classification.\fn{If 
$\varepsilon_{\tWKB}$ is insufficiently small, $\varepsilon_{\tWKB}^2$ terms from the second order 
WKB approximation become relevant, see e.g. \cite{KS}.}  
That estabished, while the ${\delta_{\sh}}\mbox{}^2\psi$ denominator of Sec 5 becomes both 
$\frac{M^2_{\mbox{\tiny Pl}}}{\hbar}\delta_{\sh}\mbox{}^2{\cal F}$ 
and $\frac{M_{\mbox{\tiny Pl}}^4}{\hbar^2}|\delta_{\sh}{\cal F}|^2$ in this Section, 
it is the latter which dominates and thus replaces 
${\delta_{\sh}}\mbox{}^2\psi$ in passing from Sec 5's approximations to this Section's.   
Thus we obtain the small quantities 
$$
\left| 
\frac{\hbar}{M_{\mbox{\scriptsize Pl}}^2}   \frac{  \langle \zeta| \delta_{\sh} \zeta\rangle}{\delta_{\sh} {\cal F}}
\right|^2 = 
\left| 
\frac{\hbar}{M_{\mbox{\scriptsize Pl}}^2}   \frac{  {\cal A}{\sh} }{\delta_{\sh}{\cal F} }
\right|^2
= \varepsilon_{\sa\sw 1}
\mbox{ } , \mbox{ }
\left| 
\frac{\hbar^2}{M^4_{\mbox{\scriptsize Pl}}} 
\frac{    \langle \lfloor\delta_{\sh}\zeta \rfloor| \delta_{\sh}\zeta\rangle   }{  |\delta_{\sh}{\cal F}|^2        }
\right|
= \varepsilon_{\sa\sw 2}
\mbox{ } , \mbox{ }
\left| 
\frac{\hbar^2}{M^4_{\mbox{\scriptsize Pl}}} \frac{    \delta_{\sh}\mbox{}^2  |\zeta\rangle   }{    |\delta_{\sh}{\cal F}|^2    } 
\right| 
= \varepsilon_{\sa\sw 3}
\mbox{ } , \mbox{ }
\left| 
\frac{\hbar^2}{M^4_{\mbox{\scriptsize Pl}}} \frac{    \langle\zeta |\delta_{\sh}\mbox{}^2|\zeta\rangle   }{    |\delta_{\sh}{\cal F}|^2    } 
\right| 
= \varepsilon_{\sa\sw 4}
\mbox{ } , \mbox{ }
$$
\be
\left| 
\frac{\hbar}{M^2_{\mbox{\scriptsize Pl}}} \frac{    \delta_{\sh} |\zeta \rangle \delta_{\sh}{\cal F}    }
                     {    |\delta_{\sh}{\cal F}|^2    }         
\right| = 
\left| 
\frac{\hbar}{M^2_{\mbox{\scriptsize Pl}}} \frac{    \delta_{\sh} |\zeta \rangle     }
                     {    \delta_{\sh}{\cal F}    }         
\right| = 
\varepsilon_{\sa\sw 7\scr} 
\mbox{ } , \mbox{ }
\left| 
\frac{\hbar}{M^2_{\mbox{\scriptsize Pl}}} \frac{    \langle \zeta |\delta_{\sh} |\zeta \rangle \delta_{\sh}{\cal F}    }
                     {    |\delta_{\sh}{\cal F}|^2    }         
\right|
= \varepsilon_{\sa\sw 8\scr} 
\mbox{ } . \mbox{ }
\ee 
Unlike in RPM, no linear constraint originating approximations are lost in passing to WKB regime.  
One has:  
\be
\left|
\frac{\hbar^2\langle\zeta|\nabla_{\alpha}{\delta_{\sh}}^{\alpha\beta}|\zeta\rangle}
     {M^4_{\Pl}\lfloor \nabla_{\alpha}{\cal F}\rfloor {\delta_{\sh}}^{\alpha\beta}{\cal F}}\right| 
= \varepsilon_{\sa\sw(\smom)1}
\mbox{ } , \mbox{ }
\left|
\frac{\hbar\lfloor \nabla_{\alpha}{\cal F} \rfloor \langle\zeta|{\delta_{\sh}}^{\alpha\beta}|\zeta\rangle}
     {M^2_{\Pl}\lfloor \nabla_{\alpha}{\cal F}\rfloor {\delta_{\sh}}^{\alpha\beta}{\cal F}}\right| 
= \varepsilon_{\sa\sw(\smom)2}
\mbox{ } , \mbox{ }
\left|
\frac{ \hbar \lfloor {\delta_{\sh}}^{\alpha\beta} {\cal F}\rfloor \langle \zeta | \nabla_{\alpha} |\zeta\rangle}
     {M^4_{\Pl} \lfloor \nabla_{\alpha}{\cal F}\rfloor {\delta_{\sh}}^{\alpha\beta}{\cal F}}\right| 
= \varepsilon_{\sa\sw(\smom)3}
\mbox{ } . \mbox{ }
\ee
There are also versions of the last 3 above built with a partly-averaged numerator, 
from consideration of the L-equation.

Mixed terms slightly change in form from the previous Section. I denote these now by `M'.  
They are 
\be
\left| 
\frac{16\pi}{\hbar c} 
\frac{    \lfloor \delta_{\sh}{\cal F}\rfloor |\delta_{\sh}|\zeta\rangle     }
     {    \delta_{\phi}\mbox{}^2|\zeta\rangle    } 
\right|
 = \varepsilon_{\sa\sM 7\scr}
\mbox{ } , \mbox{ }
\left| 
\frac{16\pi}{\hbar c} 
\frac{    \lfloor\delta_{\sh}{\cal F} \rfloor \langle \zeta | \delta_{\sh}| \zeta\rangle   }
     {  \delta_{\phi}\mbox{}^2|\zeta\rangle   }
\right|
= \left| 
\frac{16\pi}{\hbar c} 
\frac{    \lfloor\delta_{\sh}{\cal F} \rfloor {\cal A}_{\sh}   }
     {  \delta_{\phi}\mbox{}^2|\zeta\rangle   }
\right|
= \varepsilon_{\sa\sM 8\scr} 
\mbox{ } .  
\ee
The handling of ${\cal O}$ and ${\cal E}$ is as before.

Thus, the small quantities one has to contemplate at this stage are 
$\varepsilon_{\sH\sL}, \varepsilon_{\sT}, \varepsilon_{\sV}, \varepsilon_{\sI}, 
\varepsilon_{\Delta \sm}, \varepsilon_{\sA}, \varepsilon_{\ssWKB}$, 9 $\varepsilon_{\sa\sp}$ quantities, 
2 $\varepsilon_{\sa\sm}$ quantities, 
4 $\varepsilon_{\sa\sW}$ quantities, 3 $\varepsilon_{\sL}$ quantities,  $\varepsilon_{\sd\sBO}$ and the 9 from the momentum constraint.     
The W's and M's carry their w or m precursor's H/L, connection, full-path and cross-term statuses.  
The previous Section's table is then modified by these two relabellings. 
Next note that not all of these remaining $\varepsilon$'s are independent.  
This is clear from the (slight modification of the) tabulation, which reveals what excess of 
shared numerators and denominators there are. 
This affects how one can set up a full independent set of primitive quantities in terms 
of which all remaining quantities can be expressed. 
I choose to use the very cleanly adiabatic quantity
\be
|\d h/\d\phi| = \varepsilon_{\sa 1} 
\ee
as a primitive.  
Then $\varepsilon_{\sT}$ is a derived quantity, $\varepsilon_{\sT} = 
\varepsilon_{\sH\sL}/\varepsilon_{\sa 1}^2$. 
[(\ref{epsT}) is readily rearrangeable to exhibit a $m_{\phi}/M_{\Pl}$ factor].       
I also choose to use $\varepsilon_{\sa\sw 3}$ as a primitive, regardless of which path is under 
consideration.  
Then one has the following dependencies.  
\be
\varepsilon_{\sa\sp^{\prime}2} = \varepsilon_{\sa\sw 2} \frac{\varepsilon_{\sa\sw 3}}{\varepsilon_{\sa\sp 3}}
\mbox{ } , \mbox{ }
\varepsilon_{\sa\sw 1} = \varepsilon_{\sa\sp 1} \frac{\varepsilon_{\sa\sw 3}}{\varepsilon_{\sa\sp 3}}
\mbox{ } , \mbox{ }
\varepsilon_{\sa\sw 4} = \varepsilon_{\sa\sp 4} \frac{\varepsilon_{\sa\sw 3}}{\varepsilon_{\sa\sp 3}}
\mbox{ } , \mbox{ }
\varepsilon_{\sa\sw 7\scr} = \varepsilon_{\sa\sM 7\scr} \frac{\varepsilon_{\sa\sw 3}}{\varepsilon_{\sa\sp 3}}
\mbox{ } , \mbox{ }
\varepsilon_{\sa\sw 8\scr} = \varepsilon_{\sa\sM 8\scr} \frac{\varepsilon_{\sa\sw 3}}{\varepsilon_{\sa\sp 3}} \mbox{ } . 
\ee
The below-useful $\varepsilon_{\mbox{\scriptsize pert}} = {\cal I}_{\sh\phi}/{\cal V}_{\phi}$ 
 (relating to whether the H--L interaction 
can be treated as a perturbation as regards the L-subsystem) is another dependent quantity, 
being $\varepsilon_{\sI}/\varepsilon_{\sV}$.

This leaves then as a full set of primitives $\varepsilon_{\sH\sL}, \varepsilon_{\Delta \sM}, 
\varepsilon_{\Delta \sm}, \varepsilon_{\sa}, \varepsilon_{\sa 1}, \varepsilon_{\sV}, \varepsilon_{\sI}, 
\varepsilon_{\ssWKB}, \varepsilon_{\sd\sBO},$ 8 of the 9 $\varepsilon_{\sa\sp}$ (all bar 
$\varepsilon_{\sa\sp^{\prime}2}$), the 3 $\varepsilon_{\sL}$  the 2 $\varepsilon_{\sa{\sM}\scr}, \varepsilon_{\sa\sw 2}$, 
and $\varepsilon_{\sd\smom}$
and whichever alternative path's 6 from the momentum constraint.  
That's 24 from the quadratic constraint - in direct analogy with paper I - and 7 from the momentum 
constraint, so 31 in total for nontrivial geometrodynamical theories.

%====================================================================================================
%====================================================================================================
\section{A suggested interpretation of the H- and L-equations}
%====================================================================================================
%====================================================================================================

I consider H- and L-equations for geometrodynamics along the lines of Sec I.7.

%====================================================================================================
\noindent\underline{Step 1: Approximate Hamilton--Jacobi H-equation.}
%====================================================================================================
%
%
%
The coarsest approximation for the H-equation as a provider of an approximate time standard for 
the L-equation is obtained by regarding 
$\varepsilon_{\sd\sBO}, 
\varepsilon_{\sa}, 
\varepsilon_{\sH\sL}, 
\varepsilon_{\sa\sp^{\prime} 1^{\prime}},
\varepsilon_{\sa\sp^{\prime} 2^{\prime}}, 
\varepsilon_{\sa \sW   4}, 
\varepsilon_{\sa \sW   8}$
as small,  and also assuming that the averaged counterparts of 
$\varepsilon_{\sH\sL}/\varepsilon_{\sa 1}^2$, 
$\varepsilon_{\sV}$, 
$\varepsilon_{\sI}$ are small so that
\be
{\cal O} 
= \langle \hat{h} \rangle 
= \langle {\cal H}_{\sh\phi} + \sqrt{h}\{\tilde{\cal R} + \tilde{\Lambda}\} \rangle 
= \langle {\cal H}_{\sh\phi} \rangle + \sqrt{h}\{\tilde{\cal R} + \tilde{\Lambda}\}
= \langle -\hbar^2\mbox{}_{\Phi}||\pa_{\phi}||^2 + \sqrt{h}\{{\cal V}_{\phi} + 
{\cal I}_{\sh\phi}\}\rangle 
+ \sqrt{h}\{\tilde{\cal R} + \tilde{\Lambda}\}
\ee 
reduces to $\sqrt{h}\{\tilde{\cal R} + \tilde{\Lambda}\}$.   
One thus obtains the H-background GR Hamilton--Jacobi (HJ) equation 
(see \cite{Peres, Wheeler}) 
\be
\frac{\hbar}{cM_{\Pl}^2}\mbox{}_{{\cal G}}||\delta_{\sh}{\cal W}||^2  = 
\frac{c^3M_{\Pl}^2}{\hbar}\sqrt{h}\{  \Lambda   +  {\cal R}  \} 
\mbox{ } . 
\label{HJE}
\ee
The tractability in practice of this problem improves considerably if there is only one H d.o.f. 
This is e.g. the case in the common cosmological setting in which the scale factor greatly 
dominates the dynamics.  
Formally, at least, 
\be
{\cal W}(h_{\alpha\beta}) =         \frac{M_{\Pl}^2c^2}{\hbar}  \int^{h_{\alpha\beta}} 
\sqrt{h}\mbox{ }_{\underline{\cal G}^{-1}}||\d_{\suB} {h}^{\prime}_{\alpha\beta}|| 
                                      \sqrt{\Lambda + {\cal R}(h^{\prime}_{\alpha\beta})} 
\mbox{ } .  
\ee
If this is evaluable, one should check at this stage that $\varepsilon_{{\cal W}}$ is indeed small.

%=======================================================================================================
%=======================================================================================================
\noindent\underline{Step 2: underlying implicit import of emergent time.}
%=======================================================================================================
%=======================================================================================================
%
%
%
\be
\frac{\pa{\cal W}_0}{\pa h_{\alpha\beta}} \equiv \pi^{\alpha\beta} = 
\frac{M_{\Pl}^2 c}{\hbar}\frac{1}{2\mbox{N}}
{\cal G}^{\alpha\beta\gamma\delta}\DotB h_{\gamma\delta} = 
\frac{M_{\sPl}^2 c}{2\hbar} 
{\cal G}^{\alpha\beta\gamma\delta}
\left\{
\frac{\pa h_{\alpha\beta}}{\pa T^{\WKBL}} -  \frac{1}{\mbox{N}}\pounds_{\dot{\suB}}h_{\gamma\delta}
\right\}
\mbox{ }   
\label{HJT} \mbox{ } ,  
\ee
by how momentum is defined in HJ theory, 
the momentum--velocity relation and (I.172).

%====================================================================================================
\noindent\underline{Step 3: passing to a `TDSE' for the L-system.}
%====================================================================================================
%
$$
\frac{\hbar}{c M_{\Pl}^2} 2i\hbar\int \d^3x\sqrt{h}  
\mbox{}_{{\cal G}}(\delta_{\sh}{\cal W},{\cal D}^*_{\sh}|\zeta\rangle) = 
\int \d^3x\sqrt{h} i\hbar\frac{1}{\aha} {{\cal G}}^{\alpha\beta\gamma\delta} 
\left\{
\frac{\pa h_{\gamma\delta}}{\pa\lambda} - \pounds_{\dot{\suB}}h_{\gamma\delta} 
\right\}
{\cal G}_{\alpha\beta\mu\nu}
{\cal D}_{\sh}^{*\mu\nu}|\zeta\rangle 
$$
$$
= 
\int \d^3x\sqrt{h} i\hbar\frac{1}{\aha} {\mbox{id}^{\alpha\beta}}_{\gamma\delta}  
\left\{
\frac{\pa h_{\gamma\delta}}{\pa\lambda} - \pounds_{\suB}h_{\gamma\delta} 
\right\}
{\cal D}_{\sh}^{*\gamma\delta}|\zeta\rangle
=
\int \d^3x\sqrt{h} i\hbar\frac{1}{\aha} {\mbox{id}^{\alpha\beta}}_{\gamma\delta}  
\left\{
\frac{\pa h_{\gamma\delta}}{\pa\lambda}\frac{\widetilde{{\cal D}_{\sT}^*}|\zeta\rangle}
                                            {{\cal D}h_{\gamma\delta}}   + 
\frac{\pa \mbox{B}^{\gamma}}{\pa\lambda}2 \nabla^{\delta}
\frac{\widetilde{{\cal D}_{\sh}^*}|\zeta\rangle}{{\cal D}h_{\gamma\delta}} 
\right\}
$$
$$
= \left\{ 
i\hbar\frac{\widetilde{{\cal D}_{\sT}^*}|\zeta\rangle}
                                            {{\cal D}T^{\WKBL}}
- i\hbar\frac{\widetilde{{\cal D}_{\phi}^*}|\zeta\rangle}
                                            {{\cal D}T^{\WKBL}}
- \int \d^3x \frac{\pa \mbox{B}^{\gamma}}{\pa T^{\WKBL}}
\left\{
\overline{\{\hat{\cal M}_{\phi}\}_{\gamma}}
+ 2\langle \lfloor\nabla_{\gamma}\zeta\rfloor | \hat{\pi}^{\gamma\delta}| \zeta \rangle
+ \overline{\hat{\pi}}^{\gamma\delta} \frac{\pa_{\gamma}\mbox{N}}{\mbox{N}}  |\zeta\rangle
\right\}
\right\}
$$
\be
\hspace{3in}
\times 
\left\{ 1 + O
(\varepsilon_{\sH\sL}; 
\varepsilon_{\sT}, \varepsilon_{\sV}, \varepsilon_{\sI}]  
\right\} 
\mbox{ } .
\label{swalk}
\ee
Other than already-displayed definitions, 
this working uses integration by parts in the fourth move, 
and the new definitions 
\be
\frac{\widetilde{{\cal D}_{\sT}}^*}{{\cal D} f} = \frac{\pa }{\pa f} 
- i\widetilde{{\cal A}_{\sT}} 
\mbox{ } , \mbox{ }
\frac{\widetilde{{\cal D}_{\phi}}^*}{{\cal D}\lambda} = 
\left(
\frac{\pa \phi}{\pa\lambda},\delta_{\phi}
\right) 
- i\widetilde{{\cal A}_{\phi}} 
\mbox{ } \mbox{ for } \mbox{ }
\widetilde{{\cal A}_{\sT}} = -i
\left< 
\zeta \left| \frac{\pa }{\pa\lambda} \right|\zeta
\right> 
\mbox{ } , \mbox{ }
\widetilde{{\cal A}_{\phi}} = -i
\left<
\zeta \left| (\frac{\pa \phi}{\pa\lambda},{\delta_{\phi}}) \right|\zeta
\right>
\mbox{ }  \mbox{ }
\ee
(which are dynamical connections as opposed to Berry ones).     
Also, the functional dependence in (\ref{swalk}) arises from $\aha$ depending on L-variables and hence 
one not being able to carry this exactly through $\langle \mbox{ } | \mbox{ } | \mbox{ } \rangle$, 
which is resolved by expanding.

This gives more correction terms that involve comparing various $\phi$-derivatives of the wavefunction of the universe, 
which had not been noted before,  
\be
|  \tilde{\cal A}_{\sT} |\zeta\rangle/ \delta_{\phi}\mbox{}^{2} |\zeta\rangle | = \varepsilon_{\sL 4} 
\mbox{ } , \mbox{ }
|  \dot{\phi}\delta_{\phi} |\zeta\rangle/ \delta_{\phi}\mbox{}^{2} |\zeta\rangle | = \varepsilon_{\sL 5} 
\mbox{ } , \mbox{ }
|  \tilde{\cal A}_{\phi} |\zeta\rangle/ \delta_{\phi}\mbox{}^{2} |\zeta\rangle | = \varepsilon_{\sL 6} 
\mbox{ } . \mbox{ }
\ee
The chain rule term is small if the classical adiabaticity A1 dominates over the quantum adiabatic 
ap3 term.
From this and $\varepsilon_{\sT}$ small, get the suggestive rank $\varepsilon_{\sH\sL} 
<< \varepsilon_{\sa 1}^2 << \varepsilon_{\sa 1} << \varepsilon_{\sa\sp 3}$.  
i.e. mass hierarchy outstripping some kinds of adiabaticity, and adiabatic conditions varying in size.

Thus, one obtains a `TDSE' 
\be
i\hbar \frac{\pa|\zeta\rangle}{\pa T^{\WKBL}}    =  
\int\sqrt{h}
\left\{
\overline{\hat{{\cal H}}_{\phi} + \hat{{\cal I}}_{\sh\phi}}    
\right\} 
|\zeta\rangle + \hat{R}|\zeta\rangle + 
\int   
\frac{    \pa \nB^{\gamma}    }{    \pa T^{\semi}    } 
\left\{
\overline{   \{ \hat{\cal M}_{\phi}  \}_{\gamma}   } + \hat{S}_{\gamma}
\right\} 
|\zeta\rangle 
\label{TDSEFULL}
\ee
for 
\be
\hat{{R}}^{\prime}(M_{\Pl}) = i\hbar 
\left\{
\frac{{\cal D}^*_{\phi}}{{\cal D} T^{\WKBL}} + i 
\left< 
\zeta
\left| 
\frac{\pa}{\pa T^{\WKBL}}
\right|
\zeta
\right>
\right\} 
+
\hbar^{2}\int \d^3x\sqrt{h}\hat{R}(M_{\Pl})
\ee
up to $\dot{\mbox{A}}$, $\langle \mbox{ } | \mbox{ } | \mbox{ } \rangle$ exchange 
and
\be
\hat{S}^{\gamma} = -2
\left\{
\langle
\lfloor\nabla_{\delta}\zeta\rfloor | \hat{\pi}^{\gamma\delta}|\zeta\rangle + 
\overline{\hat{\pi}}^{\gamma\delta}\frac{\pa_{\delta}\mbox{N}}{\mbox{N}} 
\right\}  
\mbox{ } .
\ee 
This term is `extra trouble' from the $\nabla$ not commuting with the overline or the N.  
The latter causes the second, foliation-dependent term to appear at a detailed enough level.

However, this again leads to two objections.   
1: there are further $\pa/\pa T^{\WKBL}$ terms in the 
$\hat{R}^{\prime}$ (which are small if $\varepsilon_{\sW 3}, \varepsilon_{\sW 4}$ are) 
so that the equation is not in general a TDSE.  
2: It is not even satisfactory as a $\phi$-equation because the $\hat{R}$ contains 
$h_{\alpha\beta}$-derivatives 
(these are small if $\varepsilon_{\sL 4}, \varepsilon_{\sL 5}, \varepsilon_{\sL 6}$ are).

(\ref{TDSEFULL}) can also be recast in a Tomonaga--Schwinger-like form.  
[It can be approximately recast as such, up to 
$\dot{\mbox{A}}$, $\langle \mbox{ } | \mbox{ } | \mbox{ } \rangle$ exchange, 
but this can be avoided by premultiplying by $\dot{\mbox{A}}$ at the start of 
a rework of calculation (\ref{swalk}).]
\be
i\hbar \frac{\pa|\zeta\rangle}{\pa \lambda}    =  
\left\{
\frac{\pa \mbox{A}}{\pa\lambda}
\left\{
\overline{\hat{\cal H}_{\phi} + \hat{\cal I}_{\sh\phi}  }    
\right\} 
+ \hat{R}^{\prime\prime}
+
\frac{    \pa \mbox{B}^{\gamma}    }{    \pa \lambda    } 
\overline{\{ \hat{\cal M}_{\phi}\}_{\gamma}  + \hat{s}_{\gamma}}  
\right\}
|\zeta\rangle 
\label{TDSEFULL2}
\ee
for
\be
\hat{R}^{\prime\prime} = i\hbar
\left\{ 
\frac{    D^*_{\sL}    }{    D\lambda   } - 
\left< 
\zeta 
\left| 
\frac{\pa}{\pa\lambda}
\right|
\zeta
\right>
\right\}
+ \hbar^2 \frac{   \pa \mbox{A}   }{    \pa\lambda   }\hat{R} 
\mbox{ } , \mbox{ } 
\hat{s}^{\gamma} = -2
\langle\lfloor\nabla_{\delta}\zeta\rfloor | \hat{\pi}^{\gamma\delta}|\zeta\rangle \mbox{ } .  
\ee
This procedure removes the most unfortunate of the previous equation's new correction terms. 
Also note upon integrating that this equation's $\lambda$'s can be considered to `cancel out', 
thus giving a temporally relational form.

%=======================================================================================================
\noindent\underline{Various forms of proposed approximate L-equations.}
%=======================================================================================================
%
%
%
In the $\phi$-equation, it is customary to neglect or miss out the terms in 
$\hat{R}^{\prime\prime}$ and $\hat{s}$, (amounting to 
$\varepsilon_{\sa\sp 1}, 
\varepsilon_{\sa\sp 3},
\varepsilon_{\sa\sp 4},
\varepsilon_{\sa\sp 5},
\varepsilon_{\sa\sp 6}$ small 
- a combination of connection term neglect 
and the typical disregard for double derivatives in calculations based on the WKB ansatz.  
Moreover, the lead chroniferous cross-term has to be regarded as non-negligible for the 
timestandard in use to emerge.  
Also, some terms which prevent the wavefunction from separating into h and $\phi$ parts 
need be kept, else, having already separated out as much h as one can in Sec 5, 
$|\zeta\rangle$ would not depend on $h_{\alpha\beta}$, so a zero factor would be contained in 
the term which is to become $i\hbar\frac{\pa|\zeta\rangle}{\pa T^{\WKBL}}$.  
There is more scope for this in this Paper than in Paper I, due to nontrivial 
kinetic coupling options.  
Sometimes furthermore dropping the fluctuation terms (wiping out the overbars) 
is alluded to in the literature.

%====================================================================================================
%====================================================================================================
\noindent\underline{Step 4: explicit emergent time estimate from H-equation.}
%====================================================================================================
%====================================================================================================
%
%
%
I now begin my suggestion of how to extend the abovedescribed standard working in the case of geometrodynamics.  
(\ref{HJT}) in (\ref{HJE}) gives, upon integrating,
\be
T^{\WKBH} - T^{\WKBH}(0) = 
\frac{1}{2c}\int \mbox{}_{\underline{\cal G}^{-1}}||{\d{h}_{\alpha\beta}}||/{\sqrt{\Lambda + {\cal R}}}
\{1 + O(\epsilon_{\sH\sL}; \varepsilon_{\sT}, \varepsilon_{\sV}, \varepsilon_{\sI}  ]\} 
\equiv J(h_{\alpha\beta})   \mbox{ } .  
\ee
One can in principle evaluate this to obtain an estimate 
\be
T^{\WKBH}_0 = T^{\WKBH}_0(h_{\alpha\beta})
\label{trel}   
\ee
under the approximations $\varepsilon_{\sT}$, $\varepsilon_{\sV}$ and $\varepsilon_{\sI}$ small.  
It is a function of the $h_{\alpha\beta}$.

%====================================================================================================
\noindent\underline{Step 5: inversion of estimate, giving a L-TDSE that is H-free.}
%====================================================================================================
%
%
%
Then, In the case of 1 H d.o.f. (appropriate both as regards 
$M_{\mbox{\scriptsize Planck}}^2 >> m_{\mbox{\scriptsize inflaton}}^2$ and 
the scale factor being far more significant than anisotropies or inhomogeneities in 
observationally viable cosmological models), 
one can in principle invert (\ref{trel}) at least on some intervals of the geometrodynamical motion:
\be
a = a(T_0^{\WKBH}) \mbox{ } . 
\label{tazenda}
\ee  
Quite a general setting for this\fn{This covers some field redefinitions of some nonminimal couplings 
as well as minimal coupling, different gauge choices, closed, open and flat choices, 
but neither other presentations of other nonminimally coupled fields nor 
QM operator ordering ambiguities.} is encapsulated by the Wheeler--DeWitt equation (WDE) 
\be
\hbar^2
\left\{
\pa_a\mbox{}^2 - \mbox{}_{\check{\Phi}^{-1}(a)}||\pa_{\phi}||^2
\right\}
|\Psi\rangle
+
\left\{
\check{\cal V}_a + \check{\cal Y}_{a\phi}
\right\}
\mbox{ } ,
\ee
where the checks denote unit absorption alongside division by the original a-dependent coefficient 
of ${\pa_a}^2$ and $\check{\cal Y}_{a\phi}$ is the now in general wholly a-dependent 
$\check{\cal V}_{\phi} + \check{\cal I}_{a\phi}$.
Thus one can formally eliminate the scalefactor a in favour of $T_0^{\WKBH}$ in the L-TDSE, 
allowing one to study it/approximations to it that are nevertheless coupled to the metric subsystem 
as $T_0^{\WKBH}$-dependent perturbations  of TDSE's. 
Now, $\widetilde{{\cal D}^*}_{\phi}|\zeta\rangle / {\cal D} T^{\WKBH}$ drops out 
as $T_0^{\WKBH}$ is independent of $\phi$, and other derivatives can be recast as T-derivatives.  
There are then in general both first and second 
time derivatives in the $\phi$-equation.    
Thus the previous paper's observations about 
KG like behaviour and yet more general behaviour carry over to this paper too.

Explicitly, after BO and WKB ans\"{a}tze, the H-equation is
\be
- \{\pa_a{\cal F}\}^2 + i\hbar\pa_a\mbox{}^2{\cal F} + 2i\hbar\lfloor\pa_a{\cal F}\rfloor\langle\zeta|\pa_a|\zeta\rangle + 
\hbar^2\langle\zeta|\{\pa_a\mbox{}^2 - \mbox{}_{\check{\Phi}^{-1}(a)}||\pa_{\phi}||^2\}|\zeta\rangle + 
\check{\cal V}_a + \langle\zeta|\check{\cal Y}_{a\phi}|\zeta\rangle
\ee
and the L-equation is
\be
\{ 1 - P_{\zeta}\}
\left\{ 
2i\hbar\lfloor\pa_a{\cal F}\rfloor\pa_a + 
\hbar^2\langle\zeta|\{\pa_a\mbox{}^2 - \mbox{}_{\check{\Phi}^{-1}(a)}||\pa_{\phi}||^2\} + 
+ \check{\cal Y}_{a\phi}\right\}
|\zeta\rangle \mbox{ } .  
\ee
Then the approximate H-equation is solved by 
${\cal F} = {\cal W}_0 = \int^a\sqrt{\check{\cal V}_a(a^{\prime})}\d a^{\prime}$.  
Use also that $g(a)\d a/\d T^{\sem} = p_a = \d{\cal W}_0/\d a = \sqrt{\check{V}_a}$ and then 
\be
\pa_{a} = \frac{\d T^{\sem}}{\d a}  \frac{\d}{\d T^{\sem}} = 
\frac{    g(a(T^{\sem}))    }{\sqrt{ \check{\cal V}(a(T^{\sem}))\} }      }\frac{\d}{\d T^{\sem}}
\mbox{ } , 
\ee
\be
\pa_a\mbox{}^2 = \frac{    g^2(a(T^{\sem}))    }{\sqrt{\check{\cal V}(a(T^{\sem})   }    }
\frac{\d^2}{\d T^{\sem}\mbox{}^2} +  
\left\{
\frac{g(a(T^{\sem})}{\check{\cal V}(a(T^{\sem}) }\frac{\d g(a(T^{\sem})}{\d T^{\sem}}   - 
\frac{g^2(a(T^{\sem})}{2\check{\cal V}^2(a(T^{\sem}) }
\frac{\d \check{\cal V}(a(T^{\sem})}{\d T^{\sem}}    
\right\}
\frac{\d}{\d T^{\sem}}
\mbox{ } , 
\ee
so as to obtain a $\phi$-equation in the form 
$$
\{1 - P_{\zeta}\}
\left\{
\left\{
i\hbar + \frac{      \hbar^2 g(a(T^{\sem}))      }{      \check{\cal V}^2(a(T^{\sem}))  }  
\left\{
\frac{  \d g(a(T^{\sem}))  }{  \d T^{\sem}    }  \check{\cal V}(a(T^{\sem})) - 
\frac{g(a(T^{\sem}))}{2} \frac{    \d\check{\cal V}(a(T^{\sem}))    }{    \d T^{\sem}    }
\right\}
\right\}
\frac{\d}{\d T^{\sem}}
\right.
$$
\be
\left.
+
\frac{    \hbar^2 g^2(a(T^{\sem}))    }{    \check{\cal V}(a(T^{\sem}))    }
\frac{    \d^2    }{    \d T^{\sem}\mbox{}^2    } 
- \hbar^2\mbox{}_{    \check{\Phi}^{-1}(a(T^{\sem}))    }||\pa_{\phi}||^2 + \check{\cal Y}_{a\phi}
\right\}
|\zeta\rangle
 = 0 
\mbox{ } .  
\ee

Also note that $T^{\mbox{\scriptsize emergent(LMB:L) }}$ and 
               $T^{\mbox{\scriptsize emergent(WKB:L) }}$ are the same by comparing the above and 
(\ref{hint}, \ref{mint}), so, collecting up the emergent time results in answer to B4, 
I can again form a

%========================================================================================================
\noindent\underline{Classical-semiclassical time lemma}.
%========================================================================================================
%
%
%

\noindent $T^{\mbox{\scriptsize emergent(LMB:L)}} = T^{\WKBL}  = \left\{ T^{\WKBH} = T^{\mbox{\scriptsize emergent(LMB:H) }} \right\} + 
O(\epsilon_{\sH\sL}; \epsilon_{\sT}, \epsilon_{\sV}, \epsilon_{\sI}]$.   
%

%=====================================================================================================
%=====================================================================================================
\noindent\underline{Geometrodynamics and the problems with the WKB procedure (B2).}
%=====================================================================================================
%=====================================================================================================
%
As regards the previous paper's idea of using the na\"{\i}ve Schr\"{o}dinger interpretation to test B2),  
in the geometrodynamical context this becomes the Hawking--Page technique \cite{HP86, HP88}
of computing timeless relative probabilities.  
One could carry this out e.g. with the Gibbons--Hawking--Stewart \cite{GHS} measure so as to investigate 
how probable inflation is within model classes.  
Inflation itself being defined by inequalities/regimes (sign of second derivative, slow roll condition),  
my suggestion amounts to proceeding likewise to investigate 
how probable is a semiclassical universe (which is also defined by inequalities/regimes, 
albeit more complicatedly, as delineated in this paper).  
However, Hawking and Page \cite{HP88} (see also \cite{Wald}) have pointed out severe limitations 
with this technique -- if something and its complement are both infinite, then one cannot meaningfully 
talk even of their relative probabilities.
Though, it could be that in some contexts semiclassicality is sufficiently ubiquitous or rare 
to produce a definite answer, or that the technique could be modified to incorporate 
practical limitations on observability \cite{GT06}.  
The `weave states' and `jump states' in loop quantum gravity might be such a context.  
The more meagre idea of using semiclassicality as a future boundary condition has 
also appeared in the loop quantum cosmology literature \cite{Bojowald}.

An additional possibility as regards B2) at the level of theories of gravity is that 
a more fundamental theory could cause cutoffs which justify the semiclassical approximation.  
On the other hand, use of semiclassicality could cut one off from the Planck scale and problem of time 
issues thereat \cite{Isham93}.
This would prevent their investigation), but might be able to supply  
protective guarantees in certain theoretical frameworks as regards sensible low energy physics. 
That oscillatory WKB solutions are well capable of existing only in certain regions is illustrated 
by figs 5 and 9 of \cite{+QCos} for single scalar field isotropic 
quantum cosmologies with various potentials, while \cite{CG} has an example for which the WKB 
regime does not hold for large, late-time universes.

%=======================================================================================================
%=======================================================================================================
\noindent\underline{Compilation of various proposed approximations}
%=======================================================================================================
%=======================================================================================================
%
%
%
There are many approximations in the semiclassical approach to geometrodynamics.  
Many are similar to those in the RPM, but there are a few extra ones and a few 
differences due to the different natures of the ZAM and momentum constraints. 
Again, that complicates testing the applicability of the WKB regime.

%=======================================================================================================
\noindent\underline{More on D1, D2 and B3.}
%=======================================================================================================
%
%
%
The need for back-reaction and all the features of back-reaction are quadratic constraint 
issues which carry straight over from  the toy situation of Paper I.  It is evident 
from considerations below that geometrodynamics presents more options than the RPM 
in this respect, e.g. kinetic coupling.     
The full system is of the form
\be
i\hbar <\pa/\pa T^{\WKBL}> = ||\pa_{\sh} {\cal W}||^2 + {\cal R} + \Lambda + \mbox{corrections} 
\mbox{ } , \mbox{ }
\ee
\be
i \hbar \pa /\pa T^{\WKBL}|\zeta> - i \hbar <\pa/\pa T^{\WKBL}>|\zeta> = 
\widehat{\ttH}_{\mbox{\scriptsize effective}}(\mbox{$\phi$-physics})|\zeta> 
\mbox{ } .
\ee
Some back-reaction is attainable by considering e.g. the habitual   
\be
M_{\Pl}^2\mbox{}_{{\cal G}}||\delta_{\sh}{\cal W}||^2  = \Lambda   + {\cal R}  
- \langle {\cal H}_{\sh\phi} \rangle 
\label{firstcorr}
\ee
or Datta's 
\be
M_{\Pl}^2\mbox{}_{{\cal G}}||\delta_{\sh}{\cal W}||^2  = \Lambda  + {\cal R} - {\cal E} \mbox{ }  
\label{secondcorr} 
\mbox{ } .  
\ee
Various even fuller schemes can be assembled by retaining both of the above corrections 
and/or retaining connection terms that correct the $\pa_{\sh}$'s.

An alternative is the iterative scheme proposed in Paper I.    
Once the L-equation has been approximately solved, 
use the approximate HJ H-equation to cancel O order terms off and, assuming the new double 
derivative is negligible, one gets an HJ equation for the correction: 
\be
\{ {\cal W}_{1, a} \}^2 + 2\kappa(a)\fW_{1,\sH} + \Kappa^2(a) = 0 \mbox{ },
\ee
for
\be
\kappa(a) = \sqrt{\check{\cal V}} - {i\hbar}\langle\zeta| \pa_{a} |\zeta\rangle    
\mbox{ } ,  
\ee
\be
\Kappa^2(a) = 
- \frac{i\hbar}{2\sqrt{\check{\cal V}(a)}}\frac{\d\check{\cal V}(a)}{\d a} 
- 2i\hbar \sqrt{\check{\cal V}(a)}\langle\zeta| \pa_{a} |\zeta\rangle    
+ \hbar^2\langle\zeta| 
\left\{
\mbox{}_{\check{\Phi}(a)}||\pa_{\phi}||^2  - \pa^2_{a} 
\right\} 
|\zeta\rangle  - \langle\zeta|{\cal Y}_{a\phi} |\zeta\rangle 
\mbox{ } .  
\ee
So, looking at the sign choice consistent with earlier workings, 
\be
{\cal W}_{1} = - \kappa + \sqrt{   \kappa^2 - \Kappa^2    }    
\mbox{ } .  
\ee
Correspondingly, 
\be
T^{\sem}_{1} - T^{\sem}_{1}(0) = \int^{a} 
\frac{   g(a^{\prime}) \d a^{\prime}    }
     {    \sqrt{  \check{\cal V}(a^{\prime})  } 
          - \kappa(a^{\prime}) +  \sqrt{  \kappa^2(a^{\prime}) - \Kappa^2(a^{\prime})  }     } \mbox{ } .  
\ee
Continuing the working requires explicit solution of the L-equation.  
Then everything above is a known quantity so it is an explicit H-equation.  
Now the $\kappa$ absorbs the $\d|\zeta\rangle/\d T^{\sem}$ term.  
Both $\kappa$ and $\Kappa$ contain back-reaction contributions.  
For estimates of which quantities are indeed small, see the specific example below, and future papers.

Paper I's qualitative and quantitative points as regards D2 carry over to geometrodynamics, 
which has additional issues as regards the momentum constraint being different 
and more complicated than Paper I's ZAM constraint.  
As regards answering B3, a contextual upgrade is required: one needs to identify what 
plays the role of `free particles' for the recovery of reality in the 
relativistic cosmological setting.  
Such as the microwave background photon bath would serve for this purpose, mediating 
ordinary L-transitions and nullifying energy gap incompatibilities between 
the expansion mode and the local matter d.o.f. modes.

%==========================================================================================================================================
%==========================================================================================================================================
%=====================================================================================================
\section{Specific Minisuperspace Example}
%=====================================================================================================
%==========================================================================================================================================
%==========================================================================================================================================

Consider a single metric H-coordinate variable $a$ (scale-factor) 
\be
{\cal T}_{a} = - a \dot{a}^2 \mbox{ } , \mbox{ } 
{\cal V}_{a} = a\{1 - H^2a^2\} \mbox{ } ,   
\ee
for $H^2 = \Lambda$, positive, and a single minimally-coupled scalar field matter L-coordinate 
variable $\phi$ 
\be
{\cal T}_{\phi} = a^3\dot{\phi}^2 \mbox{ } , \mbox{ } 
{\cal V}_{\phi} = {\cal V}(\phi) \mbox{ }  . 
\ee
These are kinetically-coupled (but that's equivalent to a potential coupling in a multiplied-up 
equation).  
The Hamiltonian constraint is then 
\be
{\cal H} \equiv -  \frac{\hbar}{cM_{\Pl}^2} \frac{\pi_{a}^2}{a} + \frac{\pi_{\phi}^2}{a^3} + 
a\{1 + a^2\{{\cal V}(\phi) - \tilde{H}^2\}\} = 0 \mbox{ } , 
\ee
which, upon quantizing as described in Sec 3 (corresponding to quantizing in a cosmic time 
gauge from ADM perspective), gives the WDE 
\be
\frac{\hbar^3}{cM_{\Pl}^2}\frac{\pa_{a}\mbox{}^2|\Psi\rangle}{a}  - \hbar^2\frac{{\pa_{\phi}}^2|\Psi\rangle}{a^3} + 
+ a\{1 + a^2\{{\cal V}(\phi) - \tilde{H}^2\}\}|\Psi\rangle = 0
\mbox{ } .
\ee
Then, after applying the BO ansatz, step ADE of Sec 4, and the WKB ansatz, the H-equation is 
\be
\frac{\hbar^3}{cM_{\Pl}^2}
\left\{
-\frac{        \{  {\cal F}_{,a}  \}^2}{a} + \frac{i}{a}{\cal F}_{,aa} +  
\frac{2i}{a}{\cal F}_{,a}\langle\zeta|\pa_{a}|\zeta\rangle
\right\}
 + 
\frac{\hbar^3}{a^3}\langle \zeta | \pa_{\phi}\mbox{}^2 | \zeta\rangle - a 
+ a^3\{H^2 - \langle \zeta |{\cal V} | \zeta\rangle\} = 0 
\ee
and the L-equation is 
\be
\{1 - P_{\zeta}\}
\left\{
\frac{\hbar^3}{cM_{\Pl}^2}
\left\{
\frac{2i{\cal F}_{,a}\pa_{a} + \pa_{a}\mbox{ }^2}{a}
\right\} 
- \frac{\hbar^2\pa_{\phi}\mbox{}^2}{a^3} - 
a^3{\cal V}(\phi) 
\right\}|\zeta\rangle = 0  
\mbox{ } .  
\ee
Adopt the `coarsest scheme' for the H-equation:
\be
-\frac{\hbar^3}{cM_{\Pl}^2} \frac{\{{\cal F}_{0,a}\}^2}{a} + 
\frac{c^3M_{\Pl}^2}{\hbar}a\{a^2H^2 - a\} = 0 \mbox{ } .  
\ee
This is a HJ equation, justifying the relabelling ${\cal F}_0 \longrightarrow {\cal W}_0$.  
It is formally solved by 
\be
{\cal W}_0^{\pm} = \pm \frac{c^2M_{\Pl}^2}{\hbar^2} \int \d a \mbox{ } a\sqrt{H^2a^2 - 1} \mbox{ } ,
\ee
which is capable of being both real and imaginary.  I choose the -- sign version for discussion below.
However, for large $a$, this exhibits real behaviour corresponding to oscillations and classical 
allowability.
The oscillatory motion corresponds to ${\cal W}_0$ real ($ a > 1/H = 1/{\sqrt{|\Lambda}|}$).
Additionally, the integral is doable,   
\be
{\cal W}_0 = - \frac{c^2M_{\Pl}^2}{\hbar^2}\frac{\{H^2a^2 - 1\}^{\frac{3}{2}}}{3H^2}  +  \mbox{ const } 
\mbox{ } .  
\ee
Also, 
\be
- \frac{c M_{\Pl}^2}{\hbar} a \frac{    \pa a    }{    \d T^{\sem}   } = 
- \frac{c M_{\Pl}^2}{\hbar} \frac{a\dot{a}}{\aha}  = 
\nP_{\sa} = 
{\cal W}_{0, a} = - \frac{c^2M_{\Pl}^2}{\hbar} a\sqrt{H^2a^2 - 1}a
\mbox{ } .  
\ee
So, 
\be
T^{\sem}_0 - T^{\sem}_0(0) = 
\frac{1}{c}\int^a{        \d a^{\prime}        }/{   \sqrt{H^2a^{\prime 2} - 1}     } =  
\frac{1}{cH}\mbox{arcosh}(Ha)
\label{barde}
\mbox{ } , 
\ee  
which is valid for $Ha > 1$, i.e. for the oscillatory domain.  
Or, inverting thereupon,
\be
a = \frac{1}{H}\mbox{cosh}(Hc\{T_0^{\sem} - T_0^{\sem}(0)\}) 
\mbox{ } .    
\label{protopert}
\ee

A self-consistency check possible at this level is that, using 
\be
\pa_{a} = \frac{    1    }{    c\mbox{sinh}(Hc\{T_0^{\sem} - T^{\sem}_0(0)\})    } 
\frac{\pa}{\pa T^{\sem}_0}
\mbox{ } ,
\label{L2}
\ee
\be
{\pa_a}^2 = 
\frac{1}{  c^2\mbox{sinh}^2(Hc\{T^{\sem} - T^{\sem}(0)\})}
\frac{\pa^2}{\pa T^{\sem}_0\mbox{}^2}
- \frac{H\mbox{cosh}(Hc\{T^{\sem} - T^{\sem}(0)\})}{c\mbox{sinh}^3(Hc\{ T^{\sem}_0 - T^{\sem}_0(0)\})}
\frac{\pa}{\pa T^{\sem}_0}
 \mbox{ } ,
\label{L3}
\ee
one can look at
\be
\epsilon_{\sW} = \left|
\frac{ \hbar   H^2  \{  2 \mbox{cosh}^2(Hc\{T^{\sem}_0 - T^{\sem}_0(0)\}) - 1 \}    }
     {  c^2M_{\Pl}^2  \mbox{sinh}^3(Hc\{T_0^{\sem} - T_0^{\sem}(0)\}) 
        \mbox{cosh}^2(Hc\{T^{\sem}_0 - T^{\sem}_0(0)\})   }
\right|
\ee
which is indeed small for sufficiently large times $\sim \mbox{ } e^{-3HcT^{\sem}_0}$.  
But it can be large for small times: using Taylor's theorem and regrouping, 
\be
1 \mbox{ } \tilde{>} \mbox{ } \frac{l_{\Pl}^2l_{\Lambda}}{l_{\sem}^3}
\label{bandarra}
\mbox{ } 
\ee 
(for $l_{\sem} = c\{T_{\sem} - T_{\sem}(0)\}$), 
which is the case for $l_{\sem} \mbox{ } \tilde{>} \mbox{ } 10^{-15}$m if 
$l_{\Lambda} \approx l_{\mbox{\scriptsize Hubble}}$.

%=======================================================================================================
\noindent\underline{Subsequent form of the L-equation.} 
%=======================================================================================================
%
%
%
$$
\{1 - P_{\zeta}\}
\left\{
- \left\{
2i + 
\frac{16\pi\hbar^2}{c^2M_{\Pl}^2}\frac{    H    }{    \mbox{sinh}^3(Hc\{T_0^{\sem} - T_0^{\sem}(0)\})    }
\right\}
\hbar\frac{\pa |\zeta\rangle}{\pa T_0^{\sem}} 
+
\frac{16\pi\hbar^3}{c^3M_{\Pl}^2}\frac{    H    }
     {    \mbox{sinh}^2(Hc\{T_0^{\sem} - T^{\sem}_0(0))\} \mbox{cosh}(Hc\{T_0^{\sem} - T^{\sem}_0(0))\}     }
\right.
$$
\be
\times \mbox{ } \frac{\pa^2|\zeta\rangle}{\pa T_0^{\sem}\mbox{}^2}
\left.
- \frac{H^3\hbar^2}{\mbox{cosh}^3(Hc\{T_0^{\sem} - T^{\sem}_0(0))\} } 
\frac{\pa|\zeta\rangle}{\pa\phi^2}
- 
\frac{\mbox{cosh}^3(Hc\{T_0^{\sem} - T^{\sem}_0(0))\}}{H^3}{\cal V}(\phi)|\zeta\rangle\right\}
= 0 
\mbox{ } , 
\ee
which is not at all of TDSE form unless certain further assumptions are made.
The ratio of the two non-TDSE time derivative terms to the TDSE time derivative term goes like 
\be
\frac{    \hbar^2 H^2    }{    M_{\Pl}^2 c^3    }
\frac{    1    }{    T_{\sem}^2    }/\frac{    \hbar    }{    T_{\sem}    } \sim 
\frac{    l_{\Pl}^2 l_{\Lambda}    }{    l_{\sem}^3   } \mbox{ } ,   
\ee 
which, as above, becomes significant if $l_{\sem} \mbox{ } \tilde{<} \mbox{ } 10^{-15} m$ 
for $l_{\Lambda} \approx l_{\mbox{\scriptsize Hubble}}$.  
Thus this may be an issue as regards the early universe,  
and moreover quite a long way away from the Planck scale.  
The observable part of the universe just being a small fraction (e.g. in inflationary setting) 
makes this figure larger rather than smaller (though a deSitter-like regime may cease to be a good model for patches 
vastly in excess of the observable part of the universe)
As regards improving the accuracy of estimation within the above model, 
within the WKB regime at least, more detailed estimation of where the TDSE 
picture breaks down would require knowing the $| \zeta \rangle$ (as its first and second 
$T^{\sem}$ derivatives might differ considerably in size).
A side-issue is whether the above example is typical in having a significant of this order of magnitude.

Note that the $T^{\sem}$ dependence in this quantum cosmological example is far more pervasive 
than in the previous paper's linearly coupled HO example -- all of its terms contain such a dependence.    
Thus the simple treatment of the potential perturbation was suggested in the previous paper has  
no obvious counterpart here.  
On the other hand, the previous Paper's suggestion of 
considering the TDSE-altering terms as `kinetic' perturbations may be extendible to the above minisuperspace 
model.

%=======================================================================================================
\noindent\underline{Substitution back into the H-equation.}
%=======================================================================================================
%
%
%
This gives, expanding ${\cal W} = {\cal W}_0 + {\cal W}_1$, using the ${\cal W}_0$ HJ equation to cancel off 
some terms and considering ${\cal W}_{1,aa}$ to be negligible:
\be
\{{\cal W}_{1,a}\}^2 + 2\kappa(a){\cal W}_{1,a} + \Kappa^2(a) = 0 
\mbox{ } ,
\ee
for 
\be
\kappa(a) = \frac{16\pi\hbar^2}{c}
\left\{ 
a\sqrt{H^2a^2 - 1} +  {i\hbar}\langle \zeta |\pa_{a}| \zeta \rangle
\right\} 
\mbox{ } , 
\ee
\be
\Kappa^2(a) = a^4\langle \zeta |{\cal V}| \zeta \rangle - 
\frac{\hbar^2}{a^2}\langle \zeta |\pa_{\phi}\mbox{}^2| \zeta \rangle
- \frac{16\pi\hbar^2}{c}i\left\{
2a\sqrt{H^2a^2 - 1}\langle \zeta |\pa_{a}| \zeta \rangle + \frac{2H^2a^2 - 1}{\sqrt{H^2a^2 - 1}}
\right\}
- \frac{16\pi\hbar^3}{cM_{\Pl}^2}\langle\zeta|\pa_a\mbox{}^2|\zeta\rangle \mbox{ } .  
\ee
Then 
\be
\fW_{1} = \int 
\left\{- \kappa(a) + \sqrt{   \kappa^2(a) - \Kappa^2(a)    }   
\right\}
\d a \mbox{ } . 
\ee
Correspondingly, 
\be
T^{\sem}_{1} - T^{\sem}_{1}(0) = \frac{1}{c}\int^a 
\frac{    \d a^{\prime}    }
     {    a\sqrt{  H^2a^2 - 1  } 
          - \kappa(a) +  \sqrt{  \kappa^2(a) - \Kappa^2(a)  }     } \mbox{ } .  
\ee

Further progress would involve solving the $T^{\sem}$-dependent perturbation of the TDSE.  
Then, one would have $|\zeta\rangle$ as an explicit function, and thus one could straightforwardly 
compute $\kappa$ and $\Kappa$.

%==========================================================================================================================================
%==========================================================================================================================================
%=====================================================================================================
\section{Semiclassical versus internal time approaches in geometrodynamics}
%=====================================================================================================
%==========================================================================================================================================
%==========================================================================================================================================

%=====================================================================================================
%=====================================================================================================
\noindent\underline{Misner time and York time.}
%=====================================================================================================
%=====================================================================================================
%
%
%
In tight analogy with the RPM situation with 
$t^{\mbox{\scriptsize scale}} = \frac{1}{2}\mbox{log}J$ and the dilational 
$t^{\mbox{\scriptsize Euler}} = \sum_i \P^i \cdot R_i$, 
in GR, Misner scale time \cite{M69} $T^{\mbox{\scriptsize Misner}} = \frac{1}{2}\mbox{log}h$ 
can be disvantageous to use as a timefunction since it needn't be monotonic 
(e.g. in the onset of recollapse in a closed cosmology).  
Moreover, switching the coordinate and momentum status of the conjugate quantities $\sqrt{h}$ and 
$Y \equiv \frac{2}{3}\frac{\pi^{\alpha\beta}h_{\alpha\beta}}{\sqrt{h}}$ by a canonical transformation 
 has the advantage that 
the dilational object $Y$ has guaranteed monotonicity \cite{York72}, at least in some sectors, 
by the constant mean curvature lapse-fixing equation.
A modest but sharp example of such a sector is the closed homogenous cosmologies with suitable matter:  
\be
\dot{Y} = \frac{\mbox{N}}{3h}
\left\{
\pi^{\sT}_{\alpha\beta}\pi^{\sT\alpha\beta} + \frac{\pi^2}{3} + \{ \mbox{Positive matter terms\} } 
\right\} 
\mbox{ } ,
\ee
the right-hand side of which is $N (> 0$ for non-frozenness) times a positive 
function.\fn{The superscript T denotes tracefree part.} 
In this case, it serves as a time $Y \equiv T^{\sYo}$.

%==========================================================================================================================================
\noindent\underline{York time in this Paper's specific example.}
%==========================================================================================================================================
%
%
%
$h_{\alpha\beta} = aS_{\alpha\beta}$ for $S_{\alpha\beta}$ the unit 3-sphere metric, 
$\pi^{\alpha\beta} = -a^{5/2}\dot{a}/\mbox{N}S^{\alpha\beta}$ and $\sqrt{h} = a^{3/2}$, so   
\be
T^{\sYo} = \frac{2}{3}\frac{\pi^{\alpha\beta}h_{\alpha\beta}}{\sqrt{h}} = - 2a^2\frac{\d a}{\d T^{\sem}} 
\mbox{ } .
\ee
Thus, the York--emergent time interrelation is 
\be
T^{\sYo} = - \frac{2}{H^3}\mbox{cosh}^2(H\{T^{\sem} - T^{\sem}(0)\})
                            \mbox{sinh}^2(H\{T^{\sem} - T^{\sem}(0)\}) 
\mbox{ } .
\ee
This is monotonic: 
\be
\frac{\d T^{\sYo}}{\d T^{\sem}} = \mbox{cosh}(H\{T^{\sem} - T^{\sem}(0)\})
\{2\mbox{sinh}^2(H\{T^{\sem} - T^{\sem}(0)\}) + \mbox{cosh}^2(H\{T^{\sem} - T^{\sem}(0)\})\} > 0 
\mbox{ } .   
\ee

For small $T^{\sem}$ (relative to the timescale set by $H$, $\{c\sqrt{\Lambda}\}^{-1}$), 
\be
T^{\sYo} \sim - \frac{2}{H^2}\{T^{\sem} - T^{\sem}(0) \} 
\mbox{ } ,
\ee
so the two are the same up to choice of origin and scale (including direction).  
However, for large $T^{\sem}$, 
\be
T^{\sYo} \sim - \frac{2}{H^3}\mbox{exp}(3HT^{\sem})
\ee
so the two are not always aligned.  
Internal time--WKB time non-alignment is also commented on in \cite{Zehbook} and \cite{Kiefer94short}.

%==========================================================================================================================================
\noindent\underline{York time reformulations of the Hamiltonian.}
%==========================================================================================================================================
%
%
%
Inverting $P_{\sY} \equiv \sqrt{h} = a^{3/2}$ and the formula 
\be
T^{\sYo} = 2\pi^aa \mbox{ } : 
\ee
\be
a = P_{\sY}\mbox{}^{2/3} \mbox{ } , \mbox{ } \pi^a = T^{\sYo}/2P_{\sY}\mbox{}^{2/3} \mbox{ } .  
\ee
Thus at the classical level, ${\cal H}$ becomes 
\be
{\cal H} = - \frac{    T^{\sYo}\mbox{}^{2}    }{    4P_{\sY}\mbox{}^{4/3}P_{\sY}\mbox{}^{2/3}   } + 
\frac{\pi_{\phi}^2}{P_{\sY}\mbox{}^2} + 
P_{\sY}\mbox{}^{2/3}(1 + P_{\sY}\mbox{}^{4/3}({\cal V}_{\phi} - H^2) = 0 
\ee
so 
\be
P_{\sY}\mbox{}^8 = \{T^{\sYo}\mbox{}^2 - \pi_{\phi}\mbox{}^{2} - P_{\sY}\mbox{}^3({\cal V}_{\phi} - H^2)\}^3 \mbox{ } ,
\ee
which, as a ninth order polynomial equation, 
has the obvious general problem of not being analytically soluble in general.  
That suffices to confirm that employing York time has no guarantee of producing 
an explicit TDSE.  
However, in the free case with no cosmological constant, 
\be
P_{\sY} = \{T^{\sYo}\mbox{}^2 - \pi_{\phi}\mbox{}^2\}^{3/8} \mbox{ } ,
\ee
which gives upon quantizing, 
\be
i\hbar \frac{\pa}{\pa T^{\sY}}|\Psi\rangle = 
\left\{ 
T^{\sYo 2} + \hbar^2\frac{\pa^2}{\pa\phi^2} 
\right\}
|\Psi\rangle
\mbox{ } .  
\ee
This however is a bizarre and complicated equation for such a system, 
raising questions of firstly how to handle it and secondly whether it at all gives agreement 
with standard quantization methods (which are applicable to so simple an underlying example).

%=====================================================================================================
%=====================================================================================================
\section{Conclusion}
%=====================================================================================================
%=====================================================================================================

\noindent\underline{Similarities and differences in formalism between the two papers.}
Quantum geoemetrodynamics is built around the prima facie timeless Wheeler--DeWitt equation (WDE).
I have presented a geometrically-based formulation of semiclassical geometrodynamics 
in quite some detail as regards the many potentially small quantities that arise in this scheme.  
This hinges upon the universe being in a WKB regime, 
and upon the retention and subsequent rearrangement of a cross-term, as regards 
the emergence of a semiclassical time.  
By these means, heavy (H) background physics provides a timestandard for local, light (L) 
physics subsystems to run with respect to.  
This procedure is widely said in the literature to replace the stationary WDE with a time-dependent 
Schr\"{o}dinger equation.    
The above formulation is found to share many relevant features with Paper I's for relational particle 
models (RPM's).  Many of these stem from the close analogy between each theory's 
quadratic constraint: the fixed-energy constraint of the RPM and the Hamiltonian constraint of GR 
(which is the classical precursor of the WDE).    
My finding for RPM's with 1 heavy degree of freedom, that if the manipulation which produces the 
`$i\hbar \pa/\pa T$' term that turns the quadratic constraint from a stationary equation to a 
time-dependent one is applied throughout the unapproximated quadratic constraint then further time 
derivative terms emerge, carries over to geometrodynamics with one heavy degree of freedom 
(the cosmologically-motivated scalefactor).  
Thus, although a time-dependent wave equation emerges, it is not in general a time-dependent 
Schr\"{o}dinger equation.  
I propose an iterative scheme for approaching this problem, in which an approximate heavy equation 
provides an approximate emergent semiclassical WKB timestandard for the light physics, 
which in turn contribute a correction to the heavy equation and hence to the timestandard and so on.  
I develop this further for a particular minisuperspace quantum cosmology example.
For this, the second time derivatives looks to occur 
in at least some cosmologically relevant epochs, though it does come hand in hand with the 
WKB approximation's second spatial derivatives becoming non-negligible, 
which lies outside the present paper's scope.  
Thus whether this QM-interpretationally and partial differential equation-theoretically 
important feature does play a significant result is subject to a number of further investigations.

The justification of the WKB ansatz in the first place remains a thorn in the whole of the above 
framework, as without this assumption the rearrangement by which a timefunction emerges breaks down.  
There being present many other quantities often tacitly or summarily argued to be small makes 
investigation of whether the WKB ansatz is applicable less directly addressable 
by specific toy models than one might have expected 
(in answer to e.g. \cite{B93, Isham93, Kuchar92}).  
I.e., one can only settle it case by case upon also making 
a large number of other assumptions, which reduce each case's statement to holding on only a small corner 
of the toy model's configuration space.  

It also means as regards question D3 posed in Sec I.1 that the semiclassical quantum gravity corrections in e.g. \cite{KS} at best  
apply only to small corners of the quantum cosmological configuration space.    
Whether the cosmologically relevant epochs of the universe do or do not lie in one of 
these regimes I mostly leave as an open question, 
albeit one for which my plethora of different epsilons may well guide to an answer.

Differences between the two papers mainly stem from the momentum constraint of GR being 
differential while the zero angular momentum (ZAM) constraint of RPM's is algebraic.  
This produces different and more numerous correction terms at each of the BO, WKB and 
`extraction of $i\hbar\pa/\pa T$' levels in the working.  
It also resists phase-geometrization -- at least there is no simple way in which the 
momentum constraint, if treated in parallel with the Hamiltonian constraint, can be 
cast in a Berry-like differential-geometric language (while there was no trouble in doing 
this for the ZAM constraint).

\noindent\underline{Relations between various concepts of time.}  
The alignment between the Leibniz--Mach--Barbour (LMB) and WKB times found for the RPM 
also holds for their GR counterparts.  
There is a deep structural level at which cosmic time enjoys an analogous status to Newtonian time.  
Both are preferred foliations and both are aligned with their respective theoretical framework's 
version of emergent (semi)classical LMB--WKB time.
The alignment of cosmic and LMB--WKB time merits the further comment that in cases in which homogeneity 
and (semi)classicality both apply, {\sl either} will do to pick out a unique, privileged timefunction.  
I.e., in this highly-symmetric case, the semiclassical and high-symmetry resolutions of the 
problem of time serve 
equally well.

In contrast to all these alignments, the hidden dilational times (Euler time for the RPM and 
York time for GR) stand somewhat apart.  
Both have been shown to be {\sl capable} on some (but not all) 
occasions of being aligned with their respective theoretical frameworks' notions of time 
(in this Paper's toy model, that is the case up to time origin and time scale choice for 
$T^{\sem} << 1/c\Lambda$).  
Both are only monotonic in certain sectors of their respective theories, 
while e.g. there are other sectors in which they are frozen and thus unavailable 
as a time notion [for York time, that corresponds to regions of spacetime that only possess 
zero mean curvature (maximal) slicings rather than more general constant mean curvature ones].
Thus emergent (semi)classical WKB--LMB time looks to be a more widely applicable notion.  
Also, while `everything in the universe' contributes to the emergent (semi)classical WKB--LMB 
timestandard, hidden dilational York and Euler timefunctions are not so attuned to the contents of the 
Universe, as potential terms do not directly contribute to these.

\noindent\underline{Further work.}
A long-term goal is to study inhomogeneous perturbations about homogeneous spacetimes (building 
upon e.g. \cite{HallHaw}).    
This is relevant as regards microwave background fluctuation and galaxy formation predictions 
via the inflationary mechanism.   
Given the new observational status of this subject, we should not be content with 
highly simplified calculations even if they are self-consistent, but rather build 
up our confidence of predictions within a more complete theoretical framework such as this Paper's.  
There is additionally going to be a level (or levels) of accuracy at which homogeneous cosmology's 
notion of a privileged time will conflict with quantum geometrodynamics' Problem of Time.    
This will occur somewhere within the study of small inhomogeneities, due to e.g. ambiguities 
in averaging procedures in operationally determining what the `homogeneous background' is \cite{Kras}, 
and the inequivalence of quantization on different foliations for sufficiently general GR models 
\cite{Kuchar92, Isham93, Kuchar99}. 
What are these levels of accuracy, and are they observationally attainable in the foreseeable future?

RPM and minisuperspace models are likely to be useful in disentangling various 
conceptual issues in the above program.  
Models such as those in Paper I and II are already likely to be sufficient to investigate 
whether relative geometric phase effects are to noticeably contribute to quantum cosmology,   
while this Paper's minisuperspace work permits as an extension 
the investigation of operator ordering issues.
More elaborate RPM and minisuperspace models such as d $> 1$ RPM's and anisotropic minisuperspaces 
will permit investigation of further features.
E.g. d $> 1$ RPM's would help in understanding the complicatory role of linear constraints 
in quantum cosmology.  
Moreover, including anisotropy and inhomogeneity contributions 
is likely to require the setting up of multiple (rather than just H--L) hierarchy models, 
and may well force us to have (various levels of) L as well as H dependence in the kinetic 
matrix which as pointed out in Paper I, is a substantial complicating factor.  
d $ > 1$ RPM's and anisotropic minisuperspaces are suitable for setting up such considerations.  
Were ulteriorly exactly soluble models of this type found to be available, additional checks 
would be possible at various stages within the abovedescribed program.
%
%peruse Isham--Blyth FRW, Taub--Nut, Bianchi IX, Kantowski--Sachs for any such.

\mbox{ }

%====================================================================================================
%====================================================================================================
\noindent{\bf{\large Acknowledgments}}
%====================================================================================================
%====================================================================================================

\mbox{ }

\noindent I thank Professor Don Page, 
Dr. Julian Barbour, Professor Gary Gibbons, 
Professor Malcolm MacCallum,   
Professor Niall \'{O} Murchadha and  
Professor Reza Tavakol for discussions, 
and Dr Julian Barbour also for reading the manuscript.    
Claire.  
Eve.
And Peterhouse, for funding me.

%=====================================================BIBLIOGRAPHY==========================================================================

\end{document}